\documentclass[useAMS]{mn2e}
\usepackage{graphicx}
\def\gsim{\;\lower4pt\hbox{${\buildrel\displaystyle >\over\sim}$}\;}
\def\lsim{\;\lower4pt\hbox{${\buildrel\displaystyle <\over\sim}$}\;}
\def\grls{\;\lower4pt\hbox{${\buildrel\displaystyle >\over <}$}\;}

\title[Unstable Dynamic Homologous Stellar Core Collapses]
{Three-dimensional Hydrodynamic\\
 Instabilities in Stellar Core Collapses}
\author[Yu-Qing Lou \& Biao Lian]{Yu-Qing
Lou$^{1,2,3}$\thanks{louyq@tsinghua.edu.cn(Y-QL);jamesmolly001@gmail.com(BL)}
  and Biao Lian$^{1}$
\\
$^{1}$Department of Physics and Tsinghua Centre for Astrophysics
 (THCA), Tsinghua University, Beijing 100084, China\\
$^{2}$Department of Astronomy and Astrophysics, the University
 of Chicago, 5640 S. Ellis Ave, Chicago, IL 60637, USA\\
$^{3}$National Astronomical Observatories, Chinese Academy
 of Sciences, A20, Datun Road, Beijing 100021, China}
\begin{document}

\date{Accepted 2011 November 10.
%Year Month xx.
Received 2011 November 09;
%Year Month xx;
in original form 2011 September 03}
%Year Month xx}

\pagerange{\pageref{firstpage}--\pageref{lastpage}} \pubyear{0000}

\maketitle

\label{firstpage}

\begin{abstract}
A spherically symmetric hydrodynamic stellar core collapse
 under gravity is time-dependent and may become unstable
 once disturbed.
Subsequent nonlinear evolutions of such growth of hydrodynamic
 instabilities may lead to various physical consequences.
Specifically for a homologously collapse of stellar core
 characterized by a polytropic exponent $\Gamma=4/3$,
we examine oscillations and/or instabilities of
 three-dimensional (3D) general polytropic perturbations.
Being incompressible, the radial component of vorticity
  perturbation always grows unstably during the same
  homologous core collapse.
For compressible 3D perturbations, the polytropic index
  $\gamma$ of perturbations can differ from $\Gamma=4/3$
  of the general polytropic hydrodynamic background flow,
  where the background specific entropy is conserved
  along streamlines and can vary in radius and time.
Our model formulation here is more general than previous ones.
 The Brunt$-$V$\ddot{\rm a}$is$\ddot{\rm a}$l$\ddot{\rm a}$
  buoyancy frequency ${\cal N}$ does not vanish, allowing
  the existence of internal gravity g$^{-}-$modes and/or
  g$^{+}-$modes, related to the sign of ${\cal N}^2$
  respectively.
Eigenvalues and eigenfunctions of various oscillatory
  and unstable perturbation modes are computed given
  asymptotic boundary conditions.
%We examine oscillations and/or instabilities of three-dimensional
%  (3D) polytropic perturbations in a
%%  relativistically hot
%  homologously collapsing stellar core characterized by a
%  polytropic exponent $\Gamma=4/3$.
%We generally allow the polytropic index $\gamma$ of 3D
%  perturbations being different from $\Gamma=4/3$ of the general
%  polytropic hydrodynamic background, where the background specific
%  entropy is conserved along streamlines and thus may vary
%  in space and time self-similarly.
%For these considerably more general situations than previous
%  model formulations, the
%  Brunt$-$V$\ddot{\rm a}$is$\ddot{\rm a}$l$\ddot{\rm a}$
%  buoyancy frequency ${\cal N}$ does not vanish, allowing
%  the presence of internal gravity g$^{-}-$modes and/or
%  g$^{+}-$modes, closely tied to the sign of ${\cal N}^2$
%  respectively.
%With specified asymptotic boundary conditions, we compute
%  eigenvalues and eigenfunctions for various oscillatory
%  and unstable perturbation modes.
%The radial component of vorticity perturbation
%  is unstable in general.
As studied in several specialized cases of Goldreich \& Weber and
  of Cao \& Lou, we further confirm that acoustic p$-$modes and
  surface f$-$modes remain stable in the current more general
  situations.
%\citet{b1} and \citet{b2}.
In comparison, g$^{-}-$modes and sufficiently high radial order
  g$^{+}-$modes are unstable, leading to inevitable convective
  motions within the collapsing stellar interior; meanwhile
  sufficiently low radial order g$^{+}-$modes remain stably
  trapped in the collapsing core.
Unstable growths of 3D g$-$mode disturbances are governed
  dominantly by the angular momentum conservation
%{\color{blue} {\it
  and modified by the gas pressure restoring force.
We note in particular that unstable temporal growths of 3D
  vortical perturbations exist even when the specific entropy
  distribution becomes uniform and $\gamma=\Gamma=4/3$.
%  in contrast to the conclusion of Goldreich \& Weber
%  that no instabilities occur.
%{\bf Sure of this?} in contrast to the acoustic p$-$mode stability
%of Goldreich \& Weber. REASONS FOR MODIFICATION: the stability of
%acoustic p$-$mode is correct, instabilities occur in perturbations
%other than those p$-$modes.
Conceptually, unstable g$-$modes might bear conceivable physical
  consequences on supernova explosions, the initial kicks of nascent
  proto-neutron stars (PNSs) of as high as
  $\sim 200-500\hbox{ km s}^{-1}$ up to $\gsim 1000\hbox{ km s}^{-1}$
%  (e.g., Scheck et al. 2006 and extensive references therein),
  and break-ups of the collapsing core, while unstable growths of
  vortical perturbations can lead to fast spins of compact objects,
  3D vortical convections inside the collapsing core for possible
  magnetohydrodynamic (MHD) dynamo actions on seed magnetic fields,
  and the generation of Rossby waves further stimulated by
  gravitational wave emissions.
\end{abstract}


%\citet{b1}, GW, hereafter; \citet{b7}; \citet{b8}; \citet{b9};
%\citet{b10}; \citet{b11}; \citet{b12}; \citet{b13}; \citet{b3};
%\citet{b2}; \citet{b4} \citet{b17}; \citet{b18}; (1989a,b);
%\citet{b19}; \citet{b20}

\section{Introduction}

Over past several decades, considerable theoretical and numerical
  efforts have been devoted to examine (in)stability properties of
  core collapsing stars before or around the emergence of supernova
  (SN) rebound shock expanding inside a massive progenitor star
  (e.g. Goldreich \& Weber 1980 -- GW hereafter; Goldreich, Lai \&
  Sahrling 1996; Lai 2000; Lai \& Goldreich 2000; Blondin, Mezzacappa
  \& DeMarino 2003; Murphy, Burrows \& Heger 2004; Lou \& Wang 2006,
  2007; Burrows et al. 2006, 2007; Lou \& Cao 2008; Cao \& Lou 2009,
  2010; Lou \& Wang 2011).
A homologously collapsing relativistically hot or degenerate gas
  sphere was first studied by GW to describe the pre-SN stellar
  core collapse process, in which the gas pressure $p$ and the
  mass density $\rho$ obey the conventional polytropic equation
  of state (EoS) $p=\kappa\rho^\Gamma$ where $\Gamma=4/3$ and
  $\kappa$ are two global constants.
%{\bf
In physically realistic situations, both parameters $\Gamma$
  and $\kappa$ do not remain constant in space and time due
  to multiple complicated nuclear processes including various
  neutrino productions, electron captures etc.
  for the inner $\sim 1.5M_{\odot}$ iron core collapse according
  to numerical simulation results (e.g. Van Riper \& Lattimer
  1981; Burrows \& Lattimer 1983; Bethe 1990; Hix et al. 2003).
%    Dimmelmeier et al. 2008).
A decrease of the homologous stellar
  core mass may occur (e.g. Bethe 1990).
One may then introduce a mean value $<\Gamma>$ as
  a pressure weighted spatial average.
Meanwhile, a few numerical simulations (e.g. Van Riper \& Lattimer
  1981; Burrows \& Lattimer 1983, 1986; Bethe 1990) show that the
  specific entropy closely related to $\kappa$ does not vary much
  in time before the pre-shock stage for a pre-SN, and that the
  mean value $<\Gamma>$ does not deviate from $4/3$ significantly.
General relativistic simulations of Dimmelmeier et al.
  (2008) appear to confirm the near constancy of
  $\Gamma$ before the emergence of a bounce shock.
%During the shock bounce process of supernova, several physical
%  quantities including the entropy may experience a jump (e.g.
%  Van Riper 1982) and our formulation would no longer work.
%  }
Moreover, other numerical simulations (e.g. Bruenn 1989a,b)
  show that the homologous solution for a stellar core
  collapse represents a fair first-order approximation.
Based upon their homologous core collapse solution, GW further
  carried out a vorticity-free (i.e. a potential flow)
  three-dimensional (3D) perturbation analysis and claimed that
  no instabilities occur for such a dynamic stellar core collapse.
Polytropic homologous core collapse solutions for $\Gamma<4/3$
  have been later explored by Yahil (1983),
%\citet{b16},
  and 3D perturbation analysis on these generalized
  solutions later show that for $\Gamma>1.09$ no
  instabilities arise (e.g. Lai 2000).
We note that all these previous 3D perturbation analyses
  are actually restricted to acoustic p$-$modes as the
  Brunt$-$V$\ddot{\rm a}$is$\ddot{\rm a}$l$\ddot{\rm a}$
  buoyancy frequency ${\cal N}$ vanishes in these model
  considerations.

Lou \& Bai (2011) presented 3D isothermal perturbations
  in a self-similar isothermal nonlinear dynamic flow.
Such 3D perturbations can be cast into a self-similar form
  to describe various possible 3D dynamic flow structures
  and configurations.
In fact, this isothermal requirement can
  be relaxed to allow a more general EoS.
%\citet{b3} Several years ago,
Lou \& Cao (2008) generalized
  the self-similar or homologous core collapse solutions of
  $\Gamma=4/3$ for a more general EoS by allowing a variable
  $\kappa$ that is actually conserved along streamlines such
  that $\kappa$ becomes an arbitrary function of the independent
  self-similar variable which is a proper combination of
  radius $r$ and time $t$;
  in other words, GW solutions only represent a special
  subclass for a conventional polytropic relativistically
  hot or degenerate gas of $\Gamma=4/3$.
%According to our more general equation of state (EoS), the
%specific entropy is conserved along streamlines, that is
%\[\Big(\frac{\partial}{\partial t}+\mathbf{u}\cdot\nabla\Big)
%\Big(\frac{p}{\rho^\Gamma}\Big)=0\ ,\] {\bf Somewhat repetitive!}
%where $\mathbf{u}$ represents the radial flow velocity and the
%radial distribution of specific entropy is allowed to be fairly
%arbitrary.
Numerical simulations for the internal structure of massive
  progenitor stars and SN explosions (e.g. Bethe et al. 1979;
  Bruenn 1985; 1989a, b; Woosley, Langer \& Weaver 1993;
  Woosley, Heger \& Weaver 2002) seem to indicate that the
  specific entropy distribution is generally non-uniform
  and can vary in space and time.

%{\bf
The stellar iron core collapse, subsequent core bounce
  and shock emergence are complicated processes and
  involve several essential aspects of physics
  including various nuclear reactions,
% (producing the elements between oxygen
%and iron and the $r-$process elements),
  neutrino trapping and escape, energy transport, a
  sophisticated EoS (both at mass densities below
  and above normal nuclear density of
  $\sim 2-3\times 10^{14}\hbox{g cm}^{-3}$), general
  relativistic effects, variations of characteristic
  physical parameters and so forth (e.g. Bethe 1990
  and references therein).
Hydrodynamic instabilities due to several relevant physical
  processes may appear before or during an SN explosion due
  to the core collapse of a massive progenitor star (e.g.,
  with a mass $M\gsim 10M_{\odot}$).
The development of numerical simulation models and codes
  (mostly spherically symmetric) over five decades has
  reached various levels of sophistication.
Within the past decade in particular, the majority of
  simulation work has been multi-dimensional in nature.
%{\bf
Recent multi-dimensional numerical simulations show that
  the growth of 2D and 3D disturbances can be significant and
  may be essential for the neutrino convection/heating and thus
  the viability of the subsequent SN explosion (e.g. Dessart
  et al. 2006; Buras et al. 2006a, b; Iwakami et al. 2008;
  Nordhaus et al. 2010b).
Among others, hydrodynamics (the so-called advective-acoustic
  ``standing accretion shock instability" or
  SASI\footnote{Physically, this SASI mechanism
  involves an amplifying feedback cycle of entropy and vorticity
  perturbations, which are advected inward and generate acoustic
  waves propagating outward to distort the shock front.
%  (Cao \& Lou 2009).
Such acoustic waves grow
%around the region
  between the PNS and accretion shock, or post-shock region.},
  see e.g. Foglizzo 2001; Blondin et al. 2003; Blondin
  \& Mezzacappa 2006) and neutrino-driven convections
  can lead to 2D instabilities.
For example, it has been shown that the $l=1$ mode is
  dominant and that a bipolar sloshing of the standing
  shock wave occurs, with strong pulsational expansions
  and contractions along the symmetry axis.
These numerical simulations indicate that the growth of
  initial perturbations seems to occur only after the nascent PNS
  is formed (e.g. Dessart et al. 2006; Nordhaus et al. 2010b).
In this context, contributions of the pre-bounce core collapse
  process to the growth of unstable perturbations should not
  be ignored, which could perhaps affect an SN
  explosion directly or indirectly (e.g., by adjusting
  the process of neutrino convection and heating).
%In many numerical simulation works an artificially put in initial
%  collapse perturbation is avoided, which may lead to a loss
%  of the growth of perturbations in the core collapse phase.
In numerical simulation studies, it is not a trivial task
  to introduce self-consistent dynamic fluctuations during
  the process of core collapse phase;
this is because in many situations, such perturbation
  solutions are not known or available a priori.
%\emph{ Instead they adopt the inevitable imprecision
%  of their codes as initial perturbations.}
Often, one has to rely on the inevitable imprecision
  of numerical codes as initial perturbations.
This might miss unstable growth effects of some perturbations
  during the earlier stellar core collapse phase.
For example,
%\emph{
  Buras et al (2006a, b) studied partly
  effects of the pre-bounce perturbations and rotations
  by numerical simulations.
They add arbitrary core collapse density perturbations, which
  would be more of acoustic nature, and find that the PNS convection
  starts slightly earlier, though nothing changes significantly.
In the numerical simulation of Iwakami et al. (2008) where small
  initial radial velocity perturbations are added,
% },
  a linear growth
  phase before the SN explosion is in fact observed, during which
  the amplitudes of the initial perturbations grow nearly
  exponentially with time $t$.
%This growth phase may be traced back to implicate possible
%  instabilities during the stellar core collapse.
%\emph{
Though they start their simulations from a post-bounce
  time, the stellar core is still contracting before the final
  explosion; this observation may still be traced back to
  implicate possible g$-$mode and/or vortical instabilities
  during the early stellar core collapse phase even when the
  shock has not yet emerged.
It deems important to treat unstable perturbations with
  care during the early core collapse process.
%}
This paper is devoted to analyze the growth of 3D unstable
  perturbations during the process of the stellar core
  collapse before the core bounce.
%  may affect an SN explosion. }

In order to catch dynamic perturbation properties, we shall adopt
  in this paper a very much simplified yet still fairly challenging
  general polytropic EoS to describe a nonlinear core collapse
  of spherical symmetry and perform 3D general polytropic
  perturbation (in)stability analysis.
Bethe (1990 and extensive references therein) reviewed the
  similarity core collapse solutions (GW; Yahil \& Lattimer
  1982; Yahil 1983) which appear to capture several gross
  features of numerical simulations.
Though limited, such analytic and semi-analytic approaches
  may reveal certain physical aspects more transparently.
Likewise, we hope to derive useful hydrodynamic instability
  properties for a more general class of homologous core
  collapses and offer valuable physical insights for
  further multi-dimensional numerical simulation
  explorations, in spite of several simplifications
  and approximations involved
%  in our nontrivial `toy'
  in our hydrodynamic model formulation.
Among others, such non-spherical effects have also been shown
  highly relevant to gravitational wave emissions during a
  violent dynamic stellar collapse involving rotating iron
  cores (e.g. Saen \& Shapiro 1978; Dimmelmeier et al.
  2008 and extensive references therein).
In numerical simulations, it is also valuable to introduce
  self-consistent perturbation solutions to initiate
  well-controlled code testing and to clearly identify
  physical effects.
%  }

The 3D polytropic perturbation analysis for $\gamma=4/3$
  on these generalized dynamic collapse solutions by
%\citet{b2}
  Cao \& Lou (2009) with a self-similar evolution/distribution
  of specific entropy show that qualitatively analogous to
  stellar oscillations about a {\it static} star (e.g.
  Cowling 1941; Cox 1976; Unno et al. 1979),
%{\bf (References)},
  all kinds of perturbation modes (i.e. acoustic p$-$modes,
  surface f$-$modes and internal gravity g$-$modes) exist,
  and instabilities can always develop for certain g$-$modes
  of sufficiently high radial orders.
Cao \& Lou (2009) demonstrated that the sign of the
  Brunt$-$V$\ddot{\rm a}$is$\ddot{\rm a}$l$\ddot{\rm a}$
  buoyancy frequency squared ${\cal N}^2$, which determines
  the existence of internal g$-$modes in static stellar models,
  still holds the key for g$-$modes during the self-similar
  dynamic evolution phase of a collapsing stellar core.
Prior perturbation analyses (e.g. GW; Lai 2000; Lai \& Goldreich
  2000) were based on the collapsing stellar core model with a
  globally constant specific entropy distribution, for which the
  gas flow is conventional polytropic and thus ${\cal N}^2=0$,
  such that all g$-$modes are actually suppressed.
Therefore, GW analysis only indicates that acoustic p$-$modes and
  f$-$modes remain stable in a homologous core collapse.
%\citet{b2}
Cao \& Lou (2009) further confirmed the stability of p$-$modes and
  f$-$modes with a variable specific entropy evolution/distribution,
  but they clearly found that all g$^{-}-$modes and sufficiently
  high-order g$^{+}-$modes are unstable during the core collapse
  of a massive progenitor star.
They estimated and proposed that the instability of the lowest
  radial order $l=1$ g$-$modes can lead to initial kicks of nascent
%{\bf
PNSs with speeds of as high as $\sim 200-500\hbox{ km s}^{-1}$
  up to $\gsim 1000\hbox{ km s}^{-1}$ (e.g., Scheck et al. 2004,
  2006, Nordhaus et al. 2010a, and references therein).
%, up to $\sim 1600\hbox{ km s}^{-1}$.
Our PNS kick proposal differs from several other suggested
  mechanisms.
For example, SN explosions in binary systems do not produce
  high enough kick speeds of PNSs (e.g. Lai et al. 2001).
Hydrodynamic simulations for anisotropic mass ejection in
  an SN explosion have produced fairly low recoil velocities
  of NS (e.g. Janka \& M\"uller 1994) or started from the
  assumption that a dipolar asymmetry was present in the
  pre-collapse iron core leading to a large anisotropy of
  an SN explosion (e.g. Burrows \& Hayes 1996).
Nevertheless, the physical origin of such pre-collpase
  fluctuations is not at all clear for the moment
  (e.g. Murphy et al. 2004).
In fact, various kinds of hydrodynamic instabilities might
  be responsible for a large-scale deformation of the
  ejecta and globally aspherical SN explosions.
Our unstable $l=1$ g$-$modes during core collapse appear
  to be plausible for generating such dipolar asymmetry
  in envelope mass ejection.
Note also that perturbation analysis of Chandrasekhar (1961)
  found the highest growth rates for the $l=1, m=0$ mode for
  thermal convection within a gravitating fluid sphere.
%  }

%\citet{b4} Cao \& Lou (2010) continued to perform the 3D
%perturbation analysis in a complementary perspective.
Adopting the stellar collapse model with a constant specific
  entropy distribution as in GW, Cao \& Lou (2010) considered
  non-isentropic perturbations with a perturbation
  polytropic index $\gamma\ne\Gamma=4/3$, for which
  ${\cal N}^2$ does not vanish.
%\citet{b4}
In a complementary perspective, the results of Cao \& Lou (2010)
  also reveal instabilities of g$^{-}-$modes and sufficiently
  high radial order g$^{+}-$modes.

To reveal behaviours of g$-$modes for the most general cases,
  we perform here a 3D perturbation analysis of the collapsing
  stellar core model by allowing not only the self-similar
  profile of the specific entropy to be arbitrary, but also
  the perturbation polytropic exponent $\gamma$ being
  different from $\Gamma=4/3$ of the dynamic background.
Under such more general conditions, the
  Brunt$-$V$\ddot{\rm a}$is$\ddot{\rm a}$l$\ddot{\rm a}$
  buoyancy frequency ${\cal N}$ does not vanish in
  general and g$-$modes exist as expected.
The instability properties of g$-$modes are further explored in
  this paper, and several possible processes which may happen
  during a pre-SN stellar core collapse are discussed in
  reference to these instabilities, including the formation
  of PNSs and initial kicks of nascent PNSs.
In addition, we also search for the mechanism of unstable
  g$-$modes in the core collapsing process of a massive
  progenitor star, and it is shown that the amplification
  of g$-$mode perturbations which are mainly convective
  are mostly due to the conservation of angular momentum.

More importantly, 3D vorticity
  perturbations in a dynamically collapsing stellar core
  with a uniform specific entropy distribution and
  perturbation polytropic exponent $\gamma=\Gamma=4/3$
%which are not discussed by GW,
  are in fact unstable.
%different from those in the static stellar
%models, which are neutrally stable.}
%{\bf Be careful with these statements! }
In the more general case of a variable specific entropy
  evolution/distribution and a perturbation polytropic
  exponent $\gamma\neq\Gamma=4/3$, the radial component of
  vorticity perturbation always grows in an unstable manner.
The growth of such vorticity perturbations may lead to fast
  spins of collapsing stellar cores, vortical convections
  or circulations, as well as nonlinear Rossby waves over
  a spinning PNS and so forth.

After an introduction of background information and our
 research motivation in Section 1, we present a hydrodynamic
 description for the dynamics of spherical self-similar
 general polytropic stellar core collapse in Section 2.
Formalism for 3D general polytropic perturbations in
 a spherically symmetric dynamic stellar collapse is
 shown in Section 3.
Model analysis and results are described in Section 4.
 Conclusions are summarized in Section 5.
Details of specific mathematical analysis can be found in
 Appendices A through D for the convenience of reference.

%\section{Dynamics of spherical self-similar
%polytropic stellar core collapse}
\section{Homologous stellar core
 collapse of polytropic gas}

Fully nonlinear hydrodynamic partial differential equations
  (PDEs) include conservations of momentum and mass, the
  Poisson equation relating the gravitational potential
  $\Phi$ and the mass density $\rho$, and a general
  polytropic EoS (i.e. specific entropy conservation
  along streamlines); these PDEs are
\begin{equation}\label{eq01}
\frac{\partial \mathbf{u}}{\partial
t}+(\mathbf{u}\cdot\nabla)\mathbf{u}=-\frac{\nabla
p}{\rho}-\nabla\Phi\ ,
\end{equation}
\begin{equation}\label{eq02}
\frac{\partial\rho}{\partial t}+\nabla\cdot(\rho\mathbf{u})=0\ ,
\end{equation}
\begin{equation}\label{eq03}
\nabla^2\Phi=4\pi G\rho\ ,
\end{equation}
\begin{equation}\label{eq04}
%p=\kappa(\mathbf{r},t)\rho^{\Gamma}\ ,\qquad
\bigg(\frac{\partial}{\partial t}
+\mathbf{u}\cdot\nabla\bigg)
\ln\bigg(\frac{p}{\rho^{\Gamma}}\bigg)=0\
,
\end{equation}
where $\mathbf{u}(\mathbf{r},\ t)$, $p(\mathbf{r},\ t)$,
 $\rho(\mathbf{r},\ t)$ and $\Phi(\mathbf{r},\ t)$ are the bulk
 flow velocity, gas pressure, mass density and gravitational
 potential of a self-gravitating flow, respectively,
%{\it
 the gradient operator $\nabla$ is in terms of the 3D
 position vector $\mathbf{r}$,
%}
 $G=6.67\times 10^{-8}\ {\rm cm}^3\ {\rm g}^{-1}{\rm s}^{-2}$ is
 the gravitational constant, and $\Gamma=4/3$ is the polytropic
 exponent for a dominant Fermi gas
%{\bf
  consisting of relativistic particles
  (mainly electrons, positrons, and
  neutrinos\footnote{Neutrino interactions at sufficiently
  low densities (say a mass density $\rho\lsim\hbox{few}\times
  10^9\hbox{g cm}^{-3}$) are rare enough that the stellar matter
  may be regarded as essentially transparent to neutrinos.
For mass densities in excess of $\sim 10^{12}-10^{13}
  \hbox{g cm}^{-3}$, the neutrino-baryon scattering becomes
  strong enough and neutrinos are effectively trapped,
  moving relative to the stellar matter only by diffusion.
Neutrino-matter collisional cross sections are proportional
  to the square of the neutrino energy. } in a
  radiation field).
  %}.
%relativistically hot
%{\bf
For a progenitor stellar core collapse prior to an SN
 explosion, the stellar core temperature is of order
 of several MeV or higher.
This is indeed very hot in terms of gas temperature,
  i.e. $\sim 10^{10}-10^{11}$ K.
Nevertheless, from the perspective of central electron chemical
  potential $\mu_e\sim 300$ MeV and the neutrino chemical
  potential $\mu_{\nu}\sim 200$ MeV, the core temperature of
  a few MeV is fairly low and the relativistic Fermi-Dirac
  gas remains highly degenerate (e.g. Bethe 1990).
%}

%As can be readily verified,
It is known that nonlinear hydrodynamic PDEs
  (\ref{eq01})$-$(\ref{eq04}) remain invariant
  under the following time reversal operation,
\begin{equation}
t\rightarrow -t,\qquad\mathbf{u}\rightarrow\mathbf{-u},\quad\
\rho\rightarrow\rho,\quad\ p\rightarrow p,\quad\
\Phi\rightarrow\Phi\ ,
\end{equation}
such that a time-dependent hydrodynamic solution for a
  stellar core collapse may be reversed to effectively
  describe a dynamic outflow or expansion.

In reference to
%\citet{b2}
 Cao \& Lou (2009), we first derive in the following
 self-similar dynamic collapse solutions of spherical
 symmetry and introduce the time-dependent spatial
 scale factor $a(t)$ as
\begin{equation}\label{scale}
a(t)=\rho_c(t)^{-1/3}\bigg(\frac{\kappa_c}{\pi G}\bigg)^{1/2}\ ,
\end{equation}
where the time-dependent $\rho_c(t)$ and the constant coefficient
  $\kappa_c$ represent respectively the values of mass density
  $\rho$ and specific entropy parameter $\kappa$ at the centre
  (hence a subscript $_c$) of a massive spherical progenitor star.
With this spatial length scale factor $a(t)$ defined by equation
  (\ref{scale}), the vector radius $\mathbf{r}$ is replaced by
  a dimensionless vector radius $\mathbf{x}=\mathbf{r}/a(t)$.
From now on, we use $\mathbf{u_0}(r,\ t)$, $\rho_0(r,\ t)$,
  $p_0(r,\ t)$ and $\Phi_0(r,\ t)$ to denote the dimensional
  unperturbed radial bulk flow velocity, mass density, gas
  pressure and gravitational potential for the self-similar
  dynamic solution of the general polytropic flow background.
The spherically symmetric background flow takes the form of
\begin{equation}\label{u0}
\mathbf{u_0}(r,\ t)=\dot{a}\mathbf{x}\ ,
\end{equation}

\begin{equation}\label{rho0}
\rho_0(r,\ t)=\bigg(\frac{\kappa_c}{\pi
G}\bigg)^{3/2}a^{-3}f^3(x)\ ,
\end{equation}

\begin{equation}\label{p0}
p_0(r,\ t)=\kappa\rho_0^{4/3}=\frac{\kappa_c^3}{(\pi G)^2}
  a^{-4}g(x)f^4(x)\ ,
\end{equation}
\begin{equation}\label{Phi0}
\Phi_0(r,\ t)=\frac{4}{3}\bigg(\frac{\kappa_c^3}
  {\pi G}\bigg)^{1/2}a^{-1}\psi(x)\ ,
\end{equation}
where $\kappa(r,\ t)$ bears the form of $\kappa_cg(x)$ with
  a constant $\kappa_c$,
  %{\bf
  and $f(x)$ is a newly
  introduced reduced function of only $x$ directly related
  the mass density $\rho_0(r,\ t)$ of the dynamic background
  flow and is to be determined by hydrodynamic equations [see
  nonlinear ODE (\ref{GLane}) below].
  %}
Substituting expressions (\ref{u0})$-$(\ref{Phi0})
  for the dynamic background flow into nonlinear
  hydrodynamic PDEs (\ref{eq01})$-$(\ref{eq04}) by
  imposing spherical symmetry and without further
  approximations, we derive a set of coupled
  nonlinear ordinary differential equations (ODEs).
It can be readily confirmed that PDEs (\ref{eq02}) and
  (\ref{eq04}) are automatically satisfied, and $g(x)$ is
  allowed to take on a fairly arbitrary form, which gives
  the freedom to specify the self-similar
  evolution/distribution of the specific entropy.
For the $g(x)=1$ case, we would have the special
  subcase of a conventional polytropic relativistically hot
  or degenerate gas with a global constant specific entropy
  parameter $\kappa=\kappa_c$ as studied previously by GW.
%\citet{b1}(GW, for short).
With a more general $g(x)$ profile for the specific
  entropy distribution/evolution, nonlinear PDE
  (\ref{eq01}) of momentum conservation leads to
\begin{equation}
-\bigg(\frac{\pi G}{\kappa_c^3}\bigg)^{1/2}a^2\ddot{a}
  =\frac{1}{x}\bigg[\frac{1}{f^3}\frac{\mbox{d}}{\mbox{d}
  x}(gf^4)+\frac{4}{3}\frac{\mbox{d}\psi}{\mbox{d}x}\bigg]\ ,
\end{equation}
where the left-hand side (LHS) depends on time $t$ only while
  the right-hand side (RHS) depends only on the independent
  self-similar variable $x\equiv |\mathbf{x}|=r/a(t)$.
We should then require both sides equal to a constant
  separation parameter $4\lambda/3$, yielding the two
  separate nonlinear ODEs below
\begin{equation}\label{time}
-\bigg(\frac{\pi G}
  {\kappa_c^3}\bigg)^{1/2}a^2\ddot{a}=\frac{4\lambda}{3}\ ,
\end{equation}
\begin{equation}\label{selfsimilar}
\frac{\mbox{d}\psi}{\mbox{d}x}=\lambda
  x-\frac{3}{4f^3}\frac{\mbox{d}}{\mbox{d}x}(gf^4)\ .
\end{equation}
ODE (\ref{time}) can be solved analytically to obtain an explicit
  temporal expression for the radial length scale factor $a(t)$
  necessary for defining the independent self-similar variable
  $x\equiv |\mathbf{x}|=r/a(t)$ by equation (\ref{scale})
  and the radial flow $\mathbf{u_0}=\dot{a}\mathbf{x}$ of
  the dynamic background by equation (\ref{u0}).
The solution for the spatial scale factor $a(t)$ from
  ODE (\ref{time}) for a homologous stellar core
  collapse (GW) is simply
\begin{equation}
a(t)=(6\lambda)^{1/3}\bigg(\frac{\kappa_c^3}
  {\pi G}\bigg)^{1/6}t^{2/3}\ ,
\end{equation}
where a positive separation constant
  $\lambda>0$ is a physical requirement.
Substituting ODE (\ref{selfsimilar}) into Poisson equation
  (\ref{eq03}) that relates the gravitational potential
  $\Phi$ and the mass density $\rho$, we readily attain
  a nonlinear ODE of $f(x)$ below
\begin{equation}\label{GLane}
\frac{1}{x^2}\frac{\mbox{d}}{\mbox{d}x}\bigg[\frac{x^2}{f^3}
\frac{\mbox{d}}{\mbox{d}x}(gf^4)\bigg]+4f^3=4\lambda\ .
\end{equation}
We can prescribe substantially different yet plausible forms
  of $g(x)$ for specific entropy evolution/distribution and
  numerically integrate nonlinear ODE (\ref{GLane}) with
  proper boundary conditions using the standard fourth-order
  Runge-Kutta integration scheme (e.g. Press et al. 1986).
For instance, it is possible to impose $f(x_b)=0$ at the moving
  boundary $x=x_b$ for a homologously collapsing stellar core
  (e.g. a PNS).
Once $f(x)$ is solved satisfying the specified boundary
  conditions, the mass density $\rho_0(r,\ t)$ and other physical
  variables characterizing the general polytropic dynamic
  background flow can all be derived accordingly [see eqns
  (\ref{u0})$-$(\ref{Phi0}) and (\ref{selfsimilar})].

For the $\lambda=0$ case, nonlinear ODE (\ref{GLane})
  reduces to the general polytropic Lane-Emden equation
  (e.g. Eddington 1926; Chandrasekhar 1939; Cao \& Lou 2009).
As an additional necessary check of the consistency, ODE
  (\ref{GLane}) reduces to equation (16) of GW precisely
  as expected for $g(x)=1$.
We would set $f(0)=1$ as a consistent normalization
  for nonlinear ODE (\ref{GLane}).
%\citet{b2}
In general, Cao \& Lou (2009) require $g'(0)=0$ and $f'(0)=0$
  for a finite sound speed at the centre, where the prime $'$
  indicates the first derivative in terms of the independent
  self-similar variable $x$.
As first shown by GW and confirmed by Lou \& Cao (2008),
%\citet{b3},
  there exists a maximum acceptable value of $\lambda>0$,
  denoted by $\lambda_M$, for $f(x)$ to vanish at a
  finite value of $x=x_b$; in the parameter regime of
  $\lambda >\lambda_M$, $f(x)$ would have a minimum but
  will not reach zero within the semi-infinite interval
  $0\leq x <+\infty$.
By numerical computations for the special case of
  $g(x)=1$, this range for sensible $\lambda$ values
  is $0\leq\lambda\leq\lambda_M=0.00654376$.

%{\bf
In our scenario, we identify this $x_b$ as the
  contracting surface of the inner homologously collapsing
  stellar core and the enclosed core mass within
  $x_b$ remains unchanged.
In earlier numerical simulations on the stellar iron core
  collapse (e.g. van Riper \& Lattimer 1981; Burrows \&
  Lattimer 1983; Dimmelmeier et al. 2008), both $\Gamma$
  and $\kappa$ change in a complicated manner along mass
  parcels as dictated by several nuclear processes.
One consequence of variable $\kappa$ and $\Gamma$ in a
  collapsing star is that the mass of the inner homologously
  collapsing core decreases during the stellar core collapse
  to form a PNS with a mass of no more than $\sim 0.5M_{\odot}$
  at the end of collapse.
In our dynamic model of stellar core collapse, $\Gamma=4/3$
  does not change and $\kappa(r,\ t)=\kappa_c g(x)$ evolves
  in a self-similar yet fairly arbitrary manner.
As the enclosed mass $M$ within $x_b$ remains constant, we
  cannot directly model the gradual mass reduction of the
  inner homologously collapsing core.
By adjusting parameters of our model, we could describe
  the evolution of inner homologously collapsing core
  with low, intermediate and high masses, separately.
We may also relax the requirement of $f(x_b)=0$ at $x=x_b$ and
  choose instead $f(x_b)=f_{\rm min}>0$ for $\lambda>\lambda_M$,
  where $f_{\rm min}>0$ is the minimum value of $f(x)>0$.
As such, this leaves freedom to match the inner homologously
  collapsing core with the outer envelope collapse.
  %}

Given the idealizations and limitations of our model
  formalism, we invoke such a self-similar
  background flow solution to represent a inner stellar core
  collapse (GW; Lou \& Cao 2008; Cao \& Lou 2009, 2010).
Our motivation here is to examine oscillations and
  instabilities of 3D general polytropic perturbations during
  such a hydrodynamic core collapse of spherical symmetry.
The results of our model analysis would reveal interesting
  physical effects to be further explored by
  multi-dimensional numerical simulation analysis
  despite of the relative simplicity of our model scenario.

%\section{3D general polytropic perturbations}
\section{General 3D disturbances in a polytropic process}

We now introduce general polytropic 3D perturbations
  to the dynamic background flow of self-similar core
  collapse described and summarized in Section 2.
Specifically, we adopt the following forms
  for general polytropic 3D perturbations,
\begin{equation}\label{pertu}
\mathbf{u}=\dot{a}(\mathbf{x}+\mathbf{v_1})\ ,
\end{equation}
\begin{equation}\label{pertrho}
\rho=\rho_0(1+f_1)\ ,
\end{equation}
\begin{equation}\label{pertp}
p=p_0(1+\beta_1)\ ,
\end{equation}
\begin{equation}\label{pertPhi}
\Phi=\frac{4}{3}\bigg(\frac{\kappa_c^3}{\pi
G}\bigg)^{1/2}a^{-1}(\psi+\psi_1)\ ,
\end{equation}
where $\rho_0(r,\ t)$ and $p_0(r,\ t)$ of the dynamic
  background flow are expressed by eqns (\ref{rho0})
  and (\ref{p0}), respectively.
%{\bf Please define notations more explicitly
%as well as their spatial and temporal dependences.}
In the background plus perturbation expressions
  (\ref{pertu})$-$(\ref{pertPhi}),
%all terms containing functional factors associated
%with subscript 1 represent 3D perturbations in the
%corresponding background physical variables.
  four dimensionless variables with subscript $1$ viz.
  $\mathbf{v_1},\ f_1,\ \beta_1,\ \psi_1$ (together with
  the pertinent multiplicative factors in front)
  %{\bf
  represent 3D perturbations superposed onto
  the corresponding hydrodynamic background variables
  $\mathbf{u_0},\ \rho_0,\ p_0,\ \Phi_0$, respectively.
  %}
For a more general consideration, we allow such 3D
  perturbations to be a general polytropic process with a
  polytropic exponent $\gamma$ being different from the
  polytropic exponent $\Gamma=4/3$ characterizing the
  spherically symmetric dynamic background for a stellar
  core collapse.
If $\gamma$ actually takes on the value for the ratio of specific
  heats, then such 3D perturbations would be adiabatic.
%{\bf
If $\gamma=\Gamma=4/3$ and with a variable $g(x)$, we come
 back to the perturbation analysis of Cao \& Lou (2009).
%}

%{\bf
Hereafter, the {\it modified} nabla operator $\nabla$
  is with respect to the spherical polar coordinates
  $(x,\ \theta,\ \varphi)$ instead of $(r,\ \theta,\
  \varphi)$ unless otherwise stated.
%}
After substituting expressions (\ref{pertu})$-$(\ref{pertPhi})
  into PDEs (\ref{eq01})$-$(\ref{eq04}) with the standard
  linearization procedure and defining a logarithmic time
\begin{equation}\label{tau}
\tau=-\ln|t|
\end{equation}
to consistently remove apparent $t$ dependence in the
  coefficients of the self-similar hydrodynamic background
  in perturbation equations, we readily obtain linearized
  equations in the following forms of
\begin{equation}\label{eq21}
\frac{\partial\mathbf{v_1}}{\partial\tau}
-\frac{\mathbf{v_1}}{3}=\frac{1}{4\lambda
f^3}\Big[\nabla(gf^4\beta_1)-f_1\nabla(gf^4)\Big]
+\frac{\nabla\psi_1}{3\lambda}\ ,
\end{equation}
\begin{equation}\label{eq22}
\nabla^2\psi_1=3f^3f_1\ ,
\end{equation}
\begin{equation}\label{eq23}
\frac{\partial f_1}{\partial
\tau}=\frac{2\nabla\cdot(f^3\mathbf{v_1})}{3f^3}\ ,
\end{equation}
\begin{equation}\label{eq24}
\frac{\partial\beta_1}{\partial
\tau}=\frac{2}{3}\bigg[\gamma(\nabla\cdot\mathbf{v_1})
+\mathbf{v_1}\cdot\bigg(4\frac{\nabla f}{f} +\frac{\nabla
g}{g}\bigg)\bigg]\ ,
\end{equation}
%{\bf equation (22) here has a sign difference as
%compared to equation (25) of GW; please check.}
%Most likely, GW has a sign typo for the perturbed
%Poisson equation; Cao Yi took note of this typo;
%Lian Biao derives from Poisson equation (3) and
%also refers to equation (27) of Cao & Lou (2009).
for 3D general polytropic perturbations in vector momentum
  equation, Poisson equation\footnote{In reference to
  equation (\ref{eq22}), it appears that perturbed Poisson
  equation (25) of GW contains a sign typo (see correct
  equation 27 of Cao \& Lou 2009).}, mass conservation, and
  the specific entropy conservation along streamlines,
  respectively.
%{\bf
In deriving equation (\ref{eq24}), we have changed
  $\Gamma$ in PDE (\ref{eq04}) into $\gamma$ for 3D
  perturbations, as mentioned in the preceding paragraph.
%}
We have also arranged the sign of $\tau$ in definition
  (\ref{tau}) such that as $t\rightarrow -0^{-}$, $\tau$
  approaches positive infinity $+\infty$.
Taking the partial derivative with respect to $\tau$ and the
  curl $\nabla\times$ operation of momentum perturbation equation
  (\ref{eq21}), and substituting linear PDEs (\ref{eq23}) and
  (\ref{eq24}) into this resulting equation, we immediately
  arrive at a vector equation in terms of $\mathbf{v_1}$
  without involving the other three perturbation variables
  $\beta_1$, $f_1$ and $\psi_1$, namely
\[2\lambda\bigg(3\frac{\partial^2}{\partial\tau^2}
-\frac{\partial}{\partial\tau}\bigg)(\nabla\times\mathbf{v_1})
=\Big[(4-3\gamma)g\nabla f+f\nabla g\Big]\times\]
\begin{equation}\label{curl}
\ \nabla(\nabla\cdot\mathbf{v_1})+3\nabla
g\times\nabla(\mathbf{v_1}\cdot\nabla f)-3\nabla
f\times\nabla(\mathbf{v_1}\cdot\nabla g)\ .
\end{equation}
%{\bf Need to discuss this equation in association with Appendix A!
%This is a growth of vortical motions. Physical implications of
%this type of vortical instabilities. No perturbations in
%gravitational potential, mass density, and pressure. }
%For $\gamma\neq 4/3$ and a constant $g$ or for $\gamma=4/3$ and a
%variable $g(x)$, the RHS of equation (\ref{curl}) does not vanish.
%For the special case of $\gamma=4/3$ and a constant $g$ (i.e.
%$\nabla g=0$) simultaneously, the RHS of equation (\ref{curl}) for
%$\mathbf{v_1}$ perturbation would vanish;
%{\bf Separation of variables requested Nov 10, 2010!}
For the separation of variables in PDE (\ref{curl}), we may write
$\mathbf{v_1}=q(\tau)\mathbf{\tilde v_1}$ where $q(\tau)$ is a
function of only $\tau$ and $\mathbf{\tilde v_1}$ depends on
$\{x,\ \theta,\ \phi\}$. It follows that
\[ \frac{4m}{3}(\nabla\times\mathbf{\tilde v_1})
=\Big[(4-3\gamma)g\nabla
f+f\nabla g\Big]\times\]
\begin{equation}\label{curl01}
\ \nabla(\nabla\cdot\mathbf{\tilde v_1})+3\nabla
g\times\nabla(\mathbf{\tilde v_1}\cdot\nabla f)-3\nabla
f\times\nabla(\mathbf{\tilde v_1}\cdot\nabla g)\
\end{equation}
and
\begin{equation}\label{curl02}
\bigg(3\frac{\partial^2}{\partial\tau^2}
-\frac{\partial}{\partial\tau}\bigg)q(\tau)
=\frac{2m}{3\lambda}q(\tau)\ ,
\end{equation}
where $m$ is the separation constant. For
  $q(\tau)\propto\exp(\eta\tau)$ in ODE (\ref{curl02}),
  we obtain $3\eta^2-\eta-2m/(3\lambda)=0$ and thus
  two roots $\eta=[1\pm (1+8m/\lambda)^{1/2}]/6$.
%{\bf
The RHS of equation (\ref{curl}) does not vanish in general,
  except for the special subcase when $\gamma=\Gamma=4/3$ and
  $g(x)=1$ (i.e. $\nabla g=0$) happen simultaneously.
  %}
%{\bf
In that special subcase, the RHS of equation (\ref{curl}) vanishes
  and the vorticity perturbation $(\nabla\times\mathbf{v_1})$ has
  an unstable temporal growth $e^{\tau/3}$ during the dynamic
  stellar core collapse, although acoustic p$-$modes and f$-$modes
  remain stable (see GW and Cao \& Lou 2009).
  %}
%I doubt whether the following words below in this paragraph are
%needed here. Perhaps it occurs too early and may confuse the referee.
We show presently with more details in Section
  4.1 that dimensionless vorticity perturbation
  $(\nabla\times\mathbf{v_1})$ unstably grows in time
  as $|t|^{-1/3}$ with a fairly arbitrary dependence
  on $(x,\ \theta,\ \phi)$.
Physically, such growing vortical motions satisfying the
  mass conservation (see Appendix A) may give rise to
  rapid core rotations and 3D circulations or convections
  inside a self-similar collapsing stellar core.
This can provide a sensible mechanism of amplifying
  seed magnetic fields by magnetohydrodynamic (MHD)
  dynamo actions in a collapsing stellar core.

We solve the 3D perturbation problem in the
  modified spherical polar coordinates $(x,\ \theta,\ \varphi)$
  and presume a sufficiently general form of flow velocity
  perturbation $\mathbf{v_1}$ as
\begin{equation}\label{eq26}
\mathbf{v_1}=\mathbf{e_r}w_{r}+\nabla_{\bot}(xw_{t})
+\nabla\times(w_{rot}\mathbf{e_r})\ ,
\end{equation}
where $\mathbf{e_r}$ stands for the unit radial vector,
  $w_r,\ w_t$ and $w_{rot}$ are three dependent variables
  characterizing the flow velocity perturbation $\mathbf{v_1}$;
  and the modified transverse gradient operator is defined by
\begin{equation}
\nabla_{\bot}\equiv\mathbf{e_{\theta}}
   \frac{\partial}{x\partial\theta}
  +\mathbf{e_{\varphi}}\frac{1}{x\sin\theta}
  \frac{\partial}{\partial\varphi}\ ,
\end{equation}
%{\bf Consistency of notations, e.g. $\nabla$ etc.?}
%without involving $x$,
where $\mathbf{e_{\theta}}$ and $\mathbf{e_{\varphi}}$
  represent two orthogonal unit vectors in the $\hat\theta$
  and $\hat\varphi$ directions, respectively.
In expression (\ref{eq26}), it is easily seen that physically
  $w_r$ stands for the radial velocity perturbation, $w_t$
  governs the tangential (i.e transverse to the radial
  direction) potential flow perturbation, and that $w_{rot}$
  is responsible for the tangential vortical (non-potential
  part) flow perturbation.
By the superposition principle, we decompose the angular
  dependence of $w_{rot}$ into the spherical harmonic function
  $Y_{l\mathbf{m}}(\theta,\ \varphi)$, characterized by a pair
  of two integral indices $l$ and $\mathbf{m}$.
It follows from expression (\ref{eq26}) that
\begin{equation}\label{Rcurl}
\mathbf{e_r}\cdot(\nabla\times\mathbf{v_1})
=-\nabla_{\bot}^2w_{rot}= l(l+1)w_{rot}/x^2\ .
\end{equation}
%As the radial component of the RHS of (\ref{curl}) for the
%curl of the momentum perturbation equation vanishes, we have
%{\it
%{\bf Email sent to Lian Biao to relate (26) and (28)! The above
%expression is incorrect!! see 26-28.pdf }
%
It is then clear that for nonzero radial vorticity perturbation,
  we must require $l\geq 1$, also expected on the intuitive ground.
It can be readily seen that the RHS of the radial component of the
  curl of momentum perturbation equation (\ref{eq21}) vanishes,
  yielding an independent differential equation of $w_{rot}$
  below in reference to expression (\ref{Rcurl}),
%}
\begin{equation}\label{wrot}
\bigg(3\frac{\partial}{\partial\tau}
-1\bigg)\nabla_{\bot}^2w_{rot}
=-\bigg(3\frac{\partial}{\partial\tau}-1\bigg)l(l+1)w_{rot}/x^2=0\ .
\end{equation}
for each spherical harmonic component $Y_{l\mathbf{m}}(\theta,\
  \varphi)$.
\begin{figure}
\includegraphics[width=84mm]{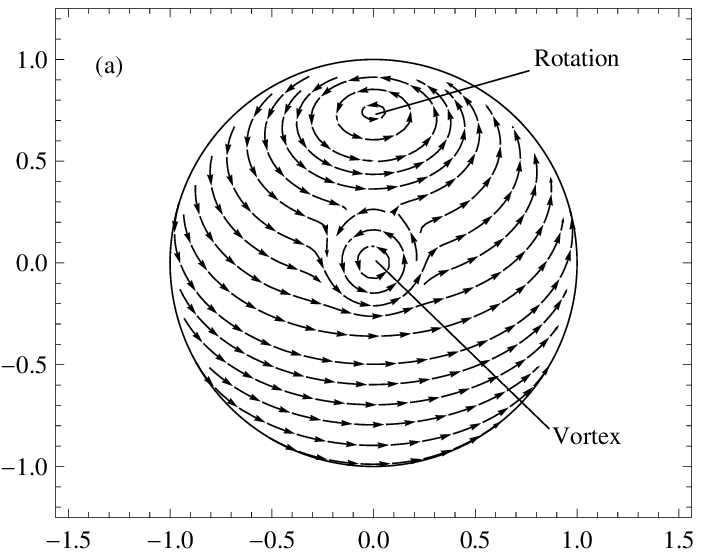}
\includegraphics[width=84mm]{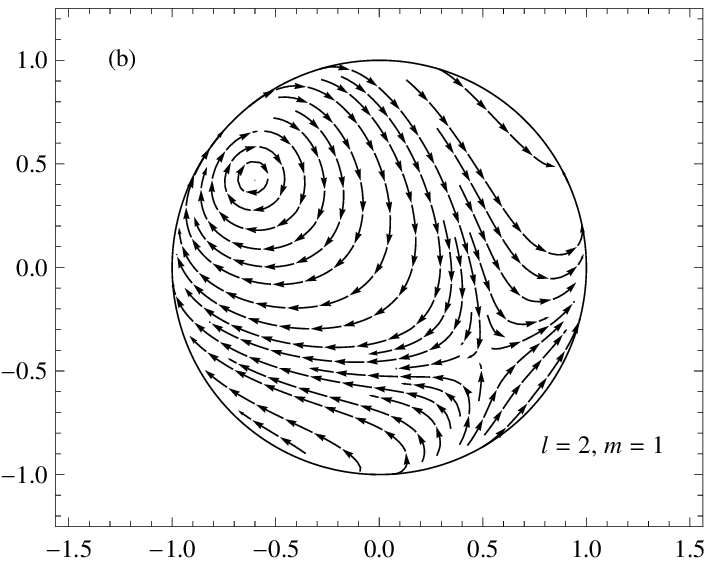}
\includegraphics[width=84mm]{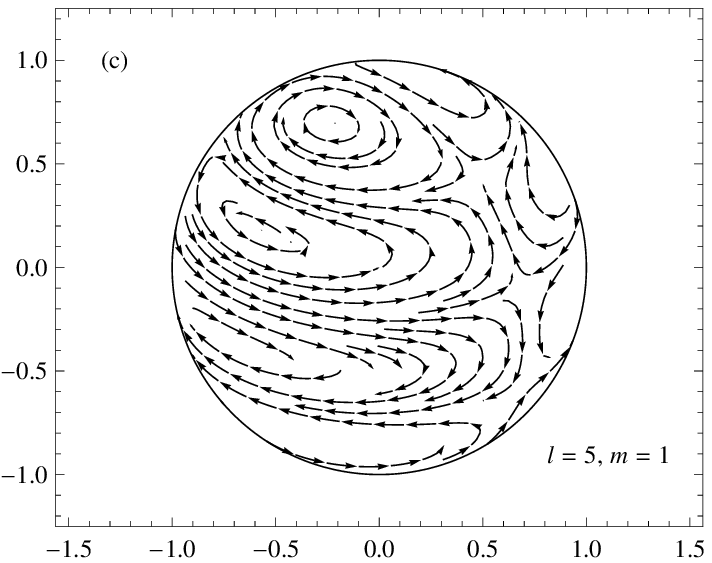}
\label{fig1} \caption{Streamlines of transverse circulations
  for three illustrating examples of tangential vortical
  motions over a spherical surface.
The patterns of such vortical perturbation motions over
  spherical surfaces can be fairly arbitrary, while the
  magnitude of velocity perturbation increases inversely
  proportional to the radius of the collapsing stellar
  core due to the conservation of angular momentum.
In the top panel, both a global spin of the collapsing
  stellar core with an embedded vortex are displayed.
For the case of $w_{rot}\propto Y_{2,1}(\theta,\ \varphi)$,
  the circulation pattern over a spherical surface is shown
  in the middle panel;
and that for the case of $w_{rot}\propto Y_{5,1}(\theta,\
  \varphi)$ is displayed in the bottom panel. }
\end{figure}
%{\it
Solving this radial vorticity perturbation equation of $w_{rot}$,
  a growing time dependence $e^{\tau/3}=|t|^{-1/3}$ for {\it all}
  spherical harmonic components of $w_{rot}$ can be readily derived.
%}{\bf The latter is extraneous because of the $\tau$ derivative?}
This temporal factor naturally shows an increase of the amplitude
  as $t\rightarrow -0^{-}$ for the radial component of vorticity
  perturbation by expression (\ref{Rcurl}).
A non-vanishing $w_{rot}$ leads to a perturbation solution for
  vortical motions with fairly arbitrary angular dependence
  in ($\theta,\ \phi$) superposed onto a dynamically collapsing
  stellar core without involving the radial component of
  velocity perturbation.
%In mathematical description, $w_{rot}$ can be associated with
%  an angular component dependence $Y_{l\mathbf{m}}(\theta,\
%  \varphi)$, characterizing the spherical harmonic function by
%  a pair of two constant integral indices $l$ and $\mathbf{m}$.
These perturbation modes of $w_{rot}$ physically
  correspond to various vortical motions transverse
  to the radial direction $\mathbf{e_r}$.
%}
%{\bf
We emphasize that the only constraint on $w_{rot}$ is ODE
  (\ref{wrot}), with no other spatial constraints (in terms
  of $x$) on $w_{rot}$.
Therefore perturbation flow component patterns of $w_{rot}$
  with various combinations of $\{ l,\ \mathbf{m}\}$ are all
  allowed and can be superposed together with a set of
  proper coefficients to form fairly general or arbitrary
  circulation patterns.
Meanwhile, it is quite clear that all
  $\{ l,\ \mathbf{m}\}$ vortical modes share the same
  temporal growth factor $e^{\tau/3}$, and they may
  form rather random tangential vortical flow patterns
  according to the superposition principle.
  %}
For the particular case of $l=1$, $\mathbf{m}=0$
  as an example, we take $w_{rot}=x^2Y_{1, 0}
  (\theta,\ \varphi)=x^2\cos\theta$, and then
  equation (\ref{eq26}) gives an expression for
  the corresponding velocity perturbation as
  $\mathbf{v_1}=x\sin\theta\ \mathbf{e_{\varphi}}$;
  and this actually describes a global spin motion
  to perturb the radially collapsing stellar core.
In the top panel of Figure 1, we show a tangential vortical
  flow pattern in which both a global spin of the stellar
  core and a relatively isolated vortex coexist.
%{\it Please provide more specific info; email sent to Lian Biao. }
It is easy to understand the time dependence $e^{\tau/3}$
  here by referring to expression (\ref{pertu}).
As can be seen, this type of vortical velocity perturbation
  is given by $\mathbf{u_1}=\dot{a}\mathbf{v_1}\propto
  e^{2\tau/3}\propto a^{-1}$.
Noting also that $w_{rot}$ represents generally vortical
  perturbations (e.g. the above perturbation example of
  rotational motion) for which
  the angular momentum is conserved, it is then clear
  that the amplitude increase of this kind of vortical velocity
  perturbation results from the conservation of angular momentum
  and the rapid radial contraction of the stellar inner core size.
For such growing vortical perturbation solutions, the
  perturbations in stellar pressure, mass density and
  gravitational potential ($f_1,\ \beta_1\ $and $\psi_1$)
  all remain zero (i.e. this is a type of incompressible
  flow perturbations).

The temporal growth factor $e^{\tau/3}$ here is faster than
  the temporal amplification factor $e^{\tau/6}$ for the
  oscillatory compressive normal modes [e.g. acoustic
  p$-$modes with $\gamma=\Gamma=4/3$ and $g(x)=1$];
this amplitude increase for all oscillatory p$-$modes
  corresponds to an adiabatic amplification of sound
  waves due to compression of the core collapse
%  due to the pure effect of background core collapse
  according to GW.
In contrast, this faster $e^{\tau/3}$ growth factor
  reveals a non-oscillatory unstable amplitude increase
  of radial vorticity perturbation during the self-similar
  dynamic phase of a stellar core collapse.
By our model scenario, the fast spin of a nascent PNS or
  a stellar mass black hole (identified as the collapsed
  stellar core remnant) is most likely triggered and
  gained during such an unstable growth of $l=1$,
  $\mathbf{m}=0$ vortical perturbations about the core
  collapse and concurrent nonlinear dynamic interactions.

%{\bf
There are alternative mechanisms
%\emph{
  proposed by others
%  }
  which might also contribute to the fast spin of a nascent PNS.
Several years ago,
%\emph{
  Blondin \& Mezzacappa (2007) suggested that
%  }
  the 3D growth of $l=1$ spiral modes (see their
  figures 1 and 2) of the SASI during the post-bounce
  delay phase may significantly spin up the core matter
  behind the stalled shock.
%\emph{
We note more specifically that this type of
  SASI fluctuations is mainly of acoustic nature.
In their 3D simulations, a nascent PNS has gained a
  considerable spin velocity with final periods of
  $\sim 70$ms for moderately rotating progenitors
  and $\sim 50$ms for non-rotating progenitors.
In contrast, more recent numerical simulations (e.g.
  Iwakami et al. 2008; Rantsiou et al. 2011) seem to
  show that such $l=1$ spiral mode mechanism does
  not work so effectively and may only induce
  possible PNS spin periods of seconds.
%   }
%Such $l=1$ spiral mode of the SASI appears to robust
%  in a stalled accretion shock in three dimensions.
%However, several numerical simulations did not show
%  such a mechanism to operate efficiently (e.g.
%%   Blondin \& Mezzacappa 2007;
%  Rantsiou et al. 2011).
Blondin \& Mezzacappa (2007) invoked several simplifications
  in their 3D simulations;
they found that the nonlinear evolution of the SASI is
  dominated by low-order non-axisymmetric mode characterized
  by a spiral flow pattern beneath the accretion shock during
  the post-bounce delay phase (e.g. Blondin \& Shaw 2007;
  Fern\`andez 2010).
%According to their simulations, the final spin periods of
%  PNSs are $\sim 70$ms for moderately rotating progenitors
%  and $\sim 50$ms for non-rotating progenitors (e.g. the
%  initial spin period of the Crab pulsar was inferred to
%  be $\sim 15-16$ms by Atoyan 1999).
Strictly speaking, they did not follow collapse itself,
  but started in a steady-state post-bounce configuration.
Also, neutrino heating and cooling were neglected and the flow
  was assumed to be isentropic in Blondin \& Mezzacappa (2007).
In short, the issue of fast PNS spin remains open and should
  be explored further for vortical instabilities proposed here.
%  }

%{\bf Spin-up of a stellar mass black hole!}
%{\it
Conceptually for the dynamic core collapse of an extremely massive
  progenitor star (e.g. Lou \& Wang 2006, 2007; Hu \& Lou 2009;
  Lou \& Wang 2011), a fast spinning central black
  hole may similarly form resulting from an unstable growth of
  such $l=1$ and $\mathbf{m}=0$ vortical perturbations.
As an example of estimation, for a core collapse in which the
  stellar radius contracts by a factor of $\sim 10^{-3}$, an initial
  stellar spin angular velocity of $\sim 1-10$ cycles per day would
  lead to a spin angular velocity of $\sim 10-100$ cycle per second
  after the core collapse.
%{\bf \emph{
Such effects might have already been observed
  in some numerical simulation works, e.g. Buras et al.
  (2006b), in which an initially specified stellar
  angular velocity is significantly amplified.
This effect may considerably or even dominantly
  spin up a nascent PNS or a stellar mass black
  hole among other alternatives.
%   } }
%}  {\bf What about the consequences of other unstable vortical
%perturbation motions with different set of $l$ and $\mathbf{m}$?
%A few more specific examples would be desirable! Rossby waves
%and gravitational wave radiation; MHD dynamo processes.}{\it
Moreover, circulation speeds of any small or large vortices
  tangent to spherical surfaces within a dynamically collapsing
  stellar core will be amplified according to this $a^{-1}$ law;
in other words, cyclones on spherical layer surfaces would spin
  faster and faster during such a dynamic core collapse.
%}

%{\bf
It is conceivable that nonlinear interactions among such
  large-scale vortices within spherical layers and stellar
  core rotation would generate and sustain Rossby-type
  inertial waves (e.g. Haurwitz 1940; Lou 1987, 2000, 2001).
Gravitational wave radiation carries angular momentum and may
  further excite stellar Rossby waves to set an upper limit
  on the spin rate of young neutron stars (e.g. Andersson
  et al. 1999; Dimmelmeier et al. 2008).
If the stellar core collapse process is accompanied with seed
  magnetic fields, these fast increasing vortex perturbations
  nonlinear interactions might also induce magnetohydrodynamic
  (MHD) dynamo actions to enhance magnetic field strengths
  (e.g. Thompson \& Duncan 1993; Hu \& Lou 2009).
%}

In the following general polytropic 3D perturbation analysis, we
  thus exclude the part involving $w_{rot}$, as we have already
  shown that it leads to unstable vortical perturbation
  solutions (i.e. radial component of vorticity perturbation)
  including the growth of rotational instability in a
  homologously collapsing stellar core.
To separate the temporal component from the angular components
  of perturbation variables, we can consistently presume that
  $\psi_1$, $\beta_1$, $f_1$, $w_r$ and $w_t$ bear the
  two-dimensional (2D) angular dependence of spherical harmonics
  $Y_{l\mathbf{m}}(\theta,\ \varphi)$ and the temporal dependence
  $e^{n\tau}$ with $n$ being an additional parameter to be determined.
Combining linearized equations (\ref{eq23}), (\ref{eq24}) and the
  two angular components of momentum perturbation equation (\ref{eq21})
  to eliminate the two dependent variables $\beta_1$ and $f_1$, it is
  straightforward to derive the following linear ODEs for 3D general
  polytropic compressible perturbations, namely
\[\frac{1}{x}\frac{\mbox{d}^2}{\mbox{d}x^2}(xn\psi_1)
-\frac{l(l+1)}{x^2}(n\psi_1)-
2f^2\bigg[\frac{f}{x^2}\frac{\mbox{d}}{\mbox{d}x}(x^2w_{r})\]
\begin{equation}\label{eq30}
\qquad\qquad\qquad\qquad-\frac{l(l+1)f}{x}
 w_{t}+3f'w_{r}\bigg]=0\ ,
\end{equation}
\[\frac{4}{3}mxw_{t}-\gamma
gf\bigg[\frac{1}{x^2}\frac{\mbox{d}}{\mbox{d}x}(x^2w_{r})
-\frac{l(l+1)}{x}w_{t}\bigg]\]
\begin{equation}\label{eq31}
\qquad\qquad\qquad\quad -(4gf'+fg')w_{r}-2(n\psi_1)=0\ ,
\end{equation}
\[\frac{4}{3}m\bigg[w_{r}-\frac{\mbox{d}}{\mbox{d}x}(xw_{t})\bigg]
+[(4-3\gamma)gf'+fg']\]
\begin{equation}\label{eq32}
\qquad\quad\ \ \times\bigg[\frac{1}{x^2}
\frac{\mbox{d}}{\mbox{d}x}(x^2w_{r})
 -\frac{l(l+1)}{x}w_{t}\bigg]=0\ .
\end{equation}
In three linear perturbation ODEs (\ref{eq30})$-$(\ref{eq32})
  above, $f'$ and $g'$ are respectively the first derivatives
  of $f(x)$ and $g(x)$ with respect to $x$, and
  $m=3\lambda n(3n-1)/2$ is the eigenvalue parameter
  to be determined by proper physical requirements.
This definition of the eigenvalue $m$ agrees
  with that of Cao \& Lou (2009).
%\citet{b2}, as may be easily verified.
In fact, equation (\ref{eq32}) is simply the angular component
  of vorticity perturbation equation (\ref{curl}).
Note that we can actually regard the product $n\psi_1$ as
  one dependent perturbation variable.\footnote{The dependent
  perturbation variables $w_{r}$ and $w_{t}$ here are proportional
  to $w_r$ and $w_t$ respectively of Cao \& Lou (2009).
%\citet{b2}.
For precise notation definitions, we should identify our $w_{r}$
  and $w_{t}$ here with $-3nw_r/2$ and $-3nw_t/2$ of Cao \& Lou (2009).}
Perturbation ODEs (\ref{eq30})$-$(\ref{eq32}) are
  equivalent to vector equation (\ref{curl01}).

For physically sensible perturbations, the relevant
  boundary conditions take the following form of
\[
%\begin{equation}
n\psi_1\propto x^l\ ,\qquad w_{r}=lw_{t}\propto x^{l-1}\qquad\
\mbox{for}\ x\rightarrow 0^{+}\ ,
\]
%\end{equation}
%\begin{equation}
\[
n\psi_1\propto x^{-(l+1)}\qquad\qquad\qquad\qquad\qquad
\mbox{for}\ x\rightarrow +\infty\ ,
\]
%\end{equation}
\begin{equation}
%\[
\frac{mxw_{t}}{3}-gf'w_{r}=\frac{n\psi_1}{2}\qquad\qquad\qquad
\mbox{for}\ x=x_b\ ;\label{BCs}
\end{equation}
%\]
here $x_b$ stands for the value of $x$
  at the moving boundary of the collapsing stellar core where
  the mass density $\rho_0$ vanishes.
These boundary conditions are imposed to avoid singularity
  in perturbation solutions as $x\rightarrow 0^{+}$ and
  $x\rightarrow +\infty$.
The condition at the contracting boundary $x_b$ requires
  a zero Lagrangian pressure perturbation there.
%{\bf Actually at $x_b$?} {\bf The meaning
%of these boundary conditions?}

%\section{Results of perturbation analysis}
\section{Model analysis and results for
three-dimensional perturbations}

%\subsection{A re-examination of GW model analysis}
\subsection{A re-visit to GW model analysis and results}

We first re-visit the special conditions adopted by GW,
  in which $\gamma=\Gamma=4/3$ and $g(x)=1$ for a
  constant specific entropy.
It follows that our equation (\ref{eq32}) reduces to
\begin{equation}\label{eq33}
m\Big[w_{r}-\frac{\mbox{d}}{\mbox{d}x}(xw_{t})\Big]=0\ .
\end{equation}
For this reason, GW choose to set
  $w_{r}=\mbox{d}(xw_{t})/\mbox{d}x$ with the eigenvalue
  parameter $m\neq 0$, leading to a kind of 3D potential
  flow velocity perturbation solutions representing only
  acoustic p$-$mode and f$-$mode oscillations in a
  homologously collapsing stellar core.
This is because internal g$-$mode oscillations
  necessarily involve vorticity perturbations.
Apparently, vorticity perturbations
  were completely excluded in the model analysis of GW.

In fact, there exist other perturbation solutions by
  setting $m=0$ in equation (\ref{eq33}), corresponding to
  either $n=0$ or $n=1/3$ yet leaving the vorticity perturbation
  factor $[w_{r}-\mbox{d}(xw_{t})/\mbox{d}x]$ being fairly arbitrary
  in terms of $x$ [see also equation (\ref{curl}) with a
  vanishing RHS].
In the former case of $n=0$, all perturbations are neutrally
  stable, while the pressure, density and gravitational
  potential perturbations do not need to vanish in general.
We now focus on the latter case of $n=1/3$, because in this case
  the perturbation is amplified during the self-similar core
  collapse process and is faster than the amplitude increase
  $e^{\tau/6}$ for the oscillatory acoustic p$-$modes due to
  adiabatic amplification of sound waves by compression in
  the core collapse.
It can be demonstrated (see Appendix A for details)
%{\bf Are you sure of this? } {\bf Not enough number of equations!}
  that to make equations (\ref{eq30}) and (\ref{eq31})
%{\bf Need to check 30 and 31?}
  meet this demand for $n=1/3$, we must have
\begin{equation}\label{eq34}
\frac{f}{x^2}\frac{\mbox{d}}{\mbox{d}x}(x^2w_{r})
-\frac{l(l+1)f}{x}w_{t}+3f'w_{r}=0\ ,
\end{equation}
corresponding to mass conservation without mass density
  perturbation, i.e. the divergence of velocity
  perturbation vanishes.
%{\bf Please indicate the physical meaning of this expression.}
This can be easily confirmed by substituting equation (\ref{eq34})
  into perturbation ODEs (\ref{eq30}) and (\ref{eq31}).
%In fact, Cowling (1941) \citet{b5} reached similar conclusions
%in his discussion of stellar oscillations based on a static
%stellar model. {\bf In what sense? Is this true?}
%The so-called Cowling approximation is one of
%Emden approximations as noted by T. G. Cowling.
For these incompressible vorticity perturbation solutions, the
  pressure, density and gravitational potential are not disturbed.
However, the time dependence of $e^{\tau/3}$ for $n=1/3$, which is
  faster than the contraction amplification factor $e^{\tau/6}$
  for oscillatory acoustic p$-$modes in the dynamic background
  core collapse (see GW), indicates a relationship between these
  growing vorticity perturbation solutions and the angular
  momentum conservation
  %{\bf
  (see the discussion on the temporal factor
  $e^{\tau/3}$ in Section 3).
  %}
In fact, these perturbation solutions represent circulatory
  motions with non-vanishing vorticities, so it is natural
  that they develop according to the angular momentum
  conservation.
In summary, we are left with ODE (\ref{eq34})
  where $[w_{r}-\mbox{d}(xw_{t})/\mbox{d}x]$
  is an arbitrary function of $x$.
By equation (\ref{curl}), this describes the growth of vortical
  convective motions, including in particular components of
  vorticity perturbation transverse to the radial direction.

%{\bf\emph{
We note that the mass density is not significantly
  disturbed in convective motions or vortex circulations;
this feature also holds approximately for internal g-modes
  to be discussed further.
In Buras et al. (2006b), they introduced arbitrary initial
  density fluctuations in the pre-bounce core collapse to
  examine effects of pre-bounce perturbations, and they
  found nothing obviously different except that the PNS
  convection starts slightly earlier.
According to our analysis above, their perturbations are
  more likely of acoustic nature rather than convective
  motions and grow much slower, having relatively minor
  effects on the post-bounce phase.
%   } }

%{\it One may note that we are left with only one equation for
%determining two variables $w_r$ and $w_t$. {\bf Need to modify
%relevant statements.} This would appear strange at a first glance.
%This problem will be clarified in the later discussion on the
%relationship between these solutions and g$-$modes in more general
%cases in subsection 4.3.}
%Later we will further discuss the relationship between
%these solutions and g$-$modes in more general cases.
%{\bf Need to think more.}\texttt{}
%\begin{figure}
%\includegraphics[width=84mm]{Fig1.eps}
%\caption{An illustration of the flow pattern of a convection
%mode inside the sellar core in GW's case, in which $l=1$.
%It is clear that the %perturbation velocity does not
%vanish at the centre of the core.} \label{}
%\end{figure}
\begin{figure}
\includegraphics[width=84mm]{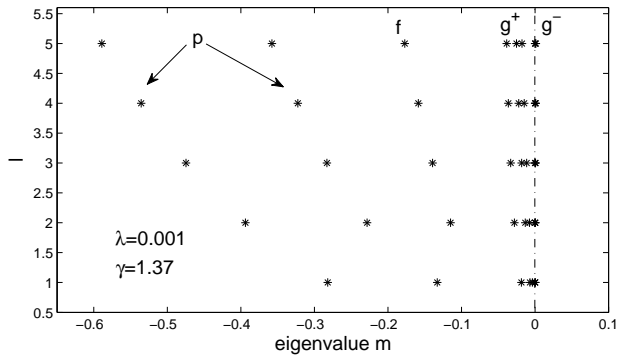}
\label{fig2}
\caption{Eigenvalues of $m$ for the case of
  $\lambda=0.001$, $\gamma=1.37>\Gamma=4/3$
  and $g(x)$ given by expression (\ref{incgx})
  as an example of illustration.
%{\it
The moving boundary of the stellar core is numerically
  determined at $x_b=6.9561$.
%{\bf Indicate this value in the figure.}
%}
In this case, both g$^{+}-$modes (only three branches are shown)
  and g$^{-}-$modes (only one branch is shown) exist.
The eigenvalues of g$^{-}-$modes (to the right of the vertical
  dash-dotted line) are quite close to zero.
The eigenvalues of the first two orders of p$-$modes are on the
  left side.
There are more branches of p$-$modes of higher orders with
  smaller eigenvalues of $m$ not shown in this figure.
The unique f$-$mode branch with $l\geq 2$ separates p$-$modes
  and g$-$modes (n.b. the $l=1$ f$-$mode does not exist).
If $\gamma$ value is higher, eigenvalues of g$^{-}-$modes
  tend to vanish.
%{\bf How to distinguish two different g-modes?}
%{\bf Please include the information of $x_b$ as
%well as in other figures.}
}
\end{figure}
\begin{figure}
\includegraphics[width=84mm]{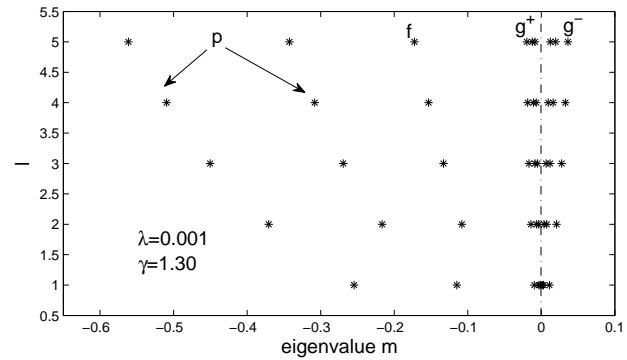}
\label{fig3}
\caption{Eigenvalues of $m$ for the case of
  $\lambda=0.001$, $\gamma=1.30<\Gamma=4/3$ and
  $g(x)$ given by expression (\ref{incgx}).
The contracting boundary of the stellar core is
  numerically determined to be at $x_b=6.9561$.
%{\bf Please also indicate this in the figure.}
Compared to those shown in Figure 2,
  the eigenvalues of $m$ in this case are
  shifted towards the positive direction.
The vertical dash-dotted line at $m=0$ separates nearby
  branches of g$^{+}-$modes (left; only three branches
  are shown) from nearby branches of g$^{-}-$modes
  (right; only three branches are shown).
When the perturbation polytropic index $\gamma$ becomes
  sufficiently small, eigenvalues of g$^{+}-$modes
  would completely disappear.
%{\bf How to distinguish two different g-modes?}
}
\end{figure}
%Among such solutions, the $l=1$ mode allows a nonzero velocity
%at the centre of the stellar core. More concretely, $m=-1,0,1$
%separately correspond to a velocity along three orthogonal
%directions at the centre. As an example, we can choose $w_{1r}$
%to take the form
%\begin{equation}
%w_{1r}=C(x-x_b)e^{\tau/3}\cos\theta\ ,
%\end{equation}
%while the corresponding expression for $w_{1t}$ is
%\begin{equation}
%w_{1t}=C\Big[\frac{3}{2}x-x_b
%+\frac{3f'}{2f}x(x-x_b)\Big]e^{\tau/3}\cos\theta\ ,
%\end{equation}
%where $C$ is a constant. The distribution of $\mathbf{w_1}$ in
%the centre section plane is illustrated in Fig. 1, from which
%we can see a motion at the centre. The $l=1$ mode probably causes
%the well known process of the "kick" of a new born neutron star,
%in which the neutron star will gain a velocity around $100 km/s$
%relative to the original gas star.

%\subsection{Numerical Explorations}
\subsection{Computational Analysis and Explorations}

In order to solve the perturbation eigenvalue problem posed here,
  we implement the procedure outlined below: first we use the
  fourth-order Runge-Kutta method (e.g. Press et al. 1986) to
  numerically integrate equation (\ref{GLane}) for obtaining
  the self-similar background function $f(x)$ which is directly
  related to the dynamic background mass density $\rho_0(r,\ t)$
  by equation (\ref{rho0}).
With this ready, we systematically discretize linear ODEs
  $(\ref{eq30})-(\ref{eq32})$ with a proper mesh to cast our
  3D perturbation problem to a matrix eigenvalue problem with
  specified boundary conditions (\ref{BCs}).
The inverse iteration method (e.g. Wilkinson 1965) is
  then employed to search for the eigenvalues of $m$
  and the corresponding eigenfunctions (see Appendix C
  for details of the numerical scheme).
%{\bf Appendix C for the inverse iteration scheme.}

We first solve the cases of globally constant specific entropy
  [i.e. $g(x)=1$] in space and time with $\gamma=\Gamma=4/3$,
  and the results of analysis agree quite well with those of
%\citet{b2},
  Cao \& Lou (2009) again confirming the errors
  of p$-$mode eigenvalues as computed by GW.
We have also checked and confirmed the results of the sample
  solutions in Cao \& Lou (2009) for several variable forms
  of $g(x)$ and $\gamma=\Gamma=4/3$ and in Cao \& Lou (2010)
  for a constant $g(x)=1$ and several $\gamma\neq\Gamma=4/3$.
%\citet{b2} and \citet{b4}.
With the assurance of a few necessary consistencies, we
  continue to apply our numerical codes to further explore
  the more general cases with a nontrivial variable $g(x)\neq 1$
  and the polytropic exponent $\gamma\ne\Gamma=4/3$.

In stellar oscillations about {\it static} stars, there exist two
  types of g$-$modes (e.g. Cox 1976) which are distinguished by
  the sign of the square of the Brunt$-$V$\ddot{\rm a}$is$\ddot{\rm
  a}$l$\ddot{\rm a}$ buoyancy frequency ${\cal N}^2$ defined by
\begin{equation}\label{BV2}
{\cal N}^2\equiv {\cal G}\bigg(\frac{\partial\ln\rho_0}{\partial
r} -\frac{1}{\gamma}\frac{\partial\ln p_0}{\partial r}\bigg)\ ,
\end{equation}
where $\rho_0$ and $p_0$ are the background gas mass
  density and thermal pressure, respectively, ${\cal G}>0$ is the
  local gravitational acceleration and $\gamma$ is the polytropic
  exponent of 3D perturbations.
For stellar oscillations, if ${\cal N}^2$ is positive everywhere,
%the eigenvalues are always negative and
  only g$^{+}-$modes occur, whereas if
  ${\cal N}^2$ is negative everywhere,
%the eigenvalues become always positive and
  only g$^{-}-$modes can be found.

For our self-similar core collapse case under consideration,
  definition (\ref{BV2}) for ${\cal N}^2$ can be extended to
  allow $\rho_0$ and $p_0$ being also time-dependent for a
  dynamic background flow.
Our numerical exploration demonstrates that if ${\cal N}^2$ is
  positive everywhere, the eigenvalues $m$ are always negative
  and only g$^{+}-$modes occur, whereas if ${\cal N}^2$ is
  negative everywhere, the eigenvalues $m$ become always
  positive and only g$^{-}-$modes can be found.
As an example of illustration, we choose the dimensionless
  function $g(x)$ for the specific entropy
  evolution/distribution in the form of
\begin{equation}\label{incgx}
g(x)=1+0.1x^2~\!\exp(-x^2/2)\ ,
\end{equation}
and take $\lambda=0.001$ for the self-similar hydrodynamic
  background.
In Figures 2 and 3, we show the eigenvalues $m$ for different
  spherical harmonic degree $l$ computed by setting
  $\gamma=1.37>4/3$ (Fig. 1) and $\gamma=1.30<4/3$ (Fig. 2),
  respectively.
In our numerical exploration, it could be easily seen that the
  eigenvalues $m$ are shifted towards the positive
  direction when $\gamma$ decreases gradually.
In Figures 2 and 3, both g$^{+}-$modes and g$^{-}-$modes exist.
However, when the value of $\gamma$ is high enough (low enough),
  eigenvalues of g$^{-}-$modes (g$^{+}-$modes) will disappear
  by examining the trends of extensive numerical results.
This can also be seen from definition (\ref{BV2}) of the
  Brunt$-$V$\ddot{\rm a}$is$\ddot{\rm a}$l$\ddot{\rm a}$
  buoyancy frequency squared ${\cal N}^2$.
Note that the eigenvalues $m=3\lambda n(3n-1)/2$ of g$-$modes
  are all fairly close to $0$, corresponding to approximately
  temporal growth factors around $e^{\tau/3}=|t|^{-1/3}$ for
  the amplitude of oscillations when $n\sim 1/3$.

%{\bf Please provide an unstable low-order
%$l=1$ g$-$mode for neutron star kicks. }

\begin{figure}
\includegraphics[width=84mm]{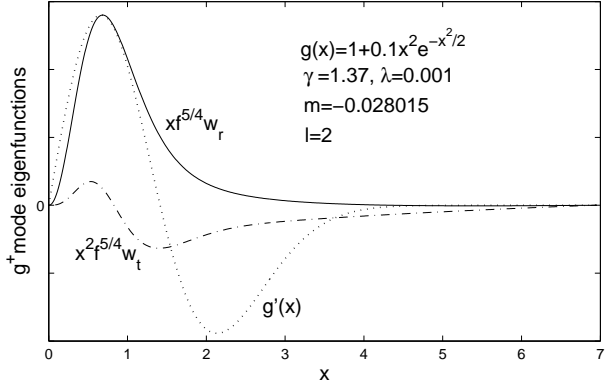}
\label{fig4}
\caption{The eigenfunctions for the lowest order
  g$^{+}-$mode with $\lambda=0.001$, $\gamma=1.37$,
  $l=2$ and $g(x)$ given by expression (\ref{incgx})
  (see Fig. 2 for sample eigenvalues).
The moving boundary of the collapsing stellar
  core is at $x_b=6.9561$.
The oscillation is primarily confined within the
  central region of the collapsing stellar core.
%{\bf Please include the information of $x_b$ and $l$.}
 }
\end{figure}
\begin{figure}
\includegraphics[width=84mm]{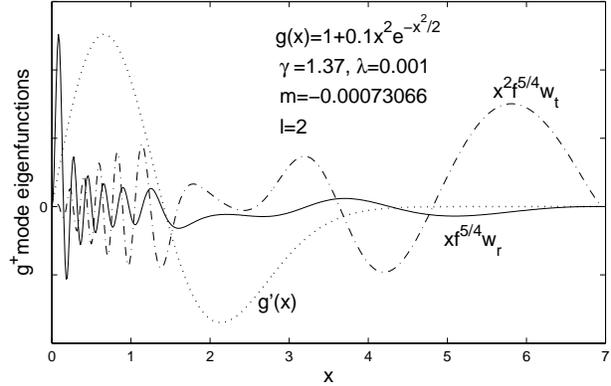}
\label{fig5} \caption{A higher order $g^{+}-$mode with
  $\lambda=0.001$, $\gamma=1.37$, $l=2$ and $g(x)$
  given by expression (\ref{incgx}).
The moving boundary of the collapsing stellar core is at
  $x_b=6.9561$. The oscillation overcomes the barrier where
  $g'(x)<0$ and approaches the outer part of the stellar core.
This appears to be a kind of wave tunneling effect. }
\end{figure}
\begin{figure}
\includegraphics[width=84mm]{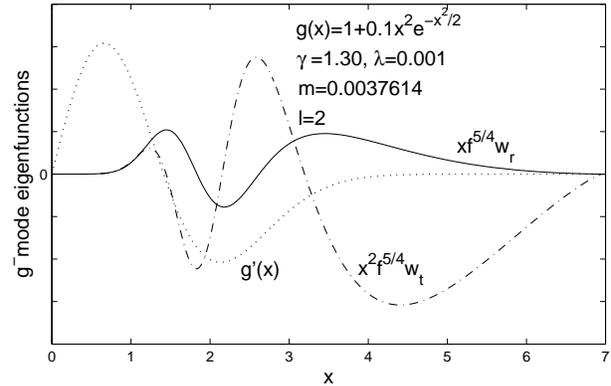}
\label{fig6} \caption{An example of g$^{-}-$mode with
 $\lambda=0.001$, $\gamma=1.30$, $l=2$ and a background
 profile $g(x)$ given by expression (\ref{incgx}).
The moving boundary of the stellar core is at $x_b=6.9561$. The
 oscillations are confined inside the region where $g'(x)<0$.}
\end{figure}

In Figure 4, we show the lowest order of g$^{+}-$modes for the
  case $\gamma=1.37$.
The solid curve stands for the scaled eigenfunction $xf^{5/4}w_r$
  (related to the radial velocity perturbation), and the dash-dotted
  curve represents the scaled eigenfunction $x^2f^{5/4}w_t$
  (related to the transverse velocity perturbation).
We also show the curve of the function $g'(x)$ by the dotted
  curve [see expression (\ref{incgx}) for the form of $g(x)$].
It can be seen that the oscillatory wave pattern of
  the eigenfunctions are mainly trapped within the
  collapsing stellar core.
However, the patterns are no longer confined inside the region
  where $g'(x)>0$, which is different from the parallel results
  in the case of $\gamma=4/3$ (see Cao \& Lou 2009).
In Figure 5, we further exhibit a higher order g$^{+}-$mode.
 It is clear that the oscillatory wave patterns are also
 mostly confined inside the
  collapsing stellar core, but the waves seem to overcome the
  barrier region (e.g. Lou 1995a,b, 1996) where $g'(x)$ is
  negative, and are amplified when approaching the outer
  boundary of the stellar core.
In the case of $\gamma\le 4/3$, however,
  this wave tunneling does not happen.
This effect seems to be due to the increase of $\gamma$ value.
Another feature of these deeply trapped g$^{+}-$modes is that
  the radial oscillation dominates, while the horizontal
  oscillation decreases inside the collapsing stellar core.

%  {\bf Sure about this last statement?}
%{\bf Need a clarification!}

Figure 6 shows one of the g$^{-}-$modes in the case of
  $\gamma=1.30$, as an example of the cases $\gamma<\Gamma=4/3$.
The oscillatory behaviours of g$^{-}-$modes are mostly confined
  within a region of $g'(x)<0$.

%{\it We further point out that for $g(x)$ close enough to constant
%$1$ and $\gamma$ close enough to $4/3$, all g$-$modes
%(g$^{+}-$modes and g$^{-}-$modes) can become unstable.
In Figure 7, we show a lowest order unstable g$^{+}-$mode for
  $\lambda=0.006$, $\gamma=4/3$, $l=1$ and $g(x)$ prescribed as
\begin{equation}\label{g4}
g(x)=1+0.001x^2\exp(-x^2/2)\ .
\end{equation}
This g$^{+}-$mode has an eigenvalue $m=-0.00014539$.
Recalling the eigenvalue definition of $m=3\lambda n(3n-1)/2$,
  we find either $n=0.3163$ or $n=0.0170$.
The former is larger than $1/6$ and corresponds
  to an unstable temporal growth.
As will be discussed in Section 4.3 presently, the $l=1$ mode
  can lead to a displacement of the very centre of the stellar core.
Such unstable core displacement along with nonlinear dynamic
  evolution may give rise to systematic movement of the
  collapsed core during the emergence of a reverse shock.
Note that this mode is mostly confined inside the collapsing
  stellar core, so the displacement of the centre of the core
  will be more pronounced than that of the outer part.
We therefore propose that the rapid increase of this unstable
  central core displacement may be ultimately responsible for
  the initial kick process of nascent PNSs or stellar mass
  black holes produced in SN remnants (see Section 4.3 and
  estimates in Cao \& Lou 2009).
%}

\begin{figure}
\includegraphics[width=84mm]{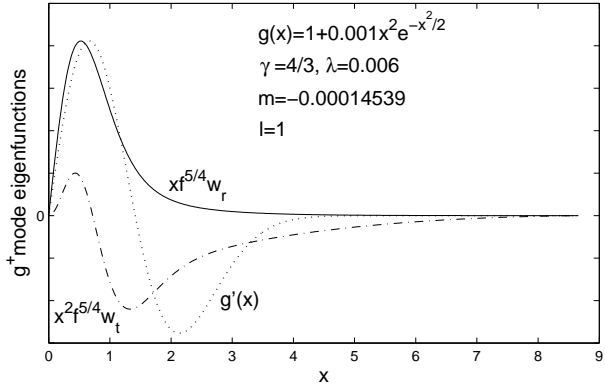}
\label{fig7}
\caption{
%{\it
An unstable lowest order g$^{+}-$mode with $\lambda=0.006$,
  $\gamma=4/3$, $l=1$ and $g(x)$ given by expression (\ref{g4}).
The contracting boundary of the collapsing stellar core is at
  $x_b=8.6598$.
 The corresponding $n$ value is either $0.3163$ or $0.0170$.
This lowest order unstable g$^{+}-$mode with $l=1$ may describe
  a kick of nascent collapsed compact object (e.g. PNS or
  stellar mass black hole left behind in a SN remnant).
%}
}
\end{figure}
Other assigned evolutions or distributions of the specific
  entropy function $g(x)$ have also been explored,
  including the following three different forms
\begin{equation}\label{g1}
 g(x)=1+\frac{0.01x^2}{0.01x^2+1}\ ,
\end{equation}
\begin{equation}\label{g2}
 g(x)=1-\frac{0.01x^2}{2(0.01x^2+1)}\ ,
\end{equation}
\begin{equation}\label{g3}
g(x)=1+0.1x^2(x-2)(x-4)\exp(-x^2/9)\ .
\end{equation}
The results of numerical solutions for 3D general polytropic
  perturbations are consistent with the earlier conclusion
  that g$^{+}-$modes are mainly confined within the regime
  with $g'(x)>0$, and that g$^{-}-$modes dominate in the
  region of $g'(x)<0$.
The specific oscillatory wave pattern may be influenced by the
  choice of the polytropic index $\gamma$ for 3D perturbations.
Regarding the last expression of $g(x)$ as defined by equation
  (\ref{g3}) or other similar expressions for which the first
  derivative $g'(x)$ crosses the value $0$ more than once, the
  region $g'(x)>0$ [$g'(x)<0$] are divided into several
  intervals in the $x$ domain, the corresponding g$^{+}-$modes
  (g$^{-}-$modes) appear to have the trend to concentrate
  inside the interval nearest to the stellar interior core.
This indicates that g$-$modes are mainly trapped deeply
  inside the central region of a collapsing stellar core.

%{\bf How about growth rates of g$-$mode instabilities? }
%{\color{red}
%{\it I suppose that it is answered in section 4.3.}

%\subsection{The mechanism for g$-$mode growths (Title?)}
%\subsection{Complementary Perspectives of Instabilities}
\subsection{Physical Nature of Perturbation Instabilities}

We have shown in Figures 2 and 3 that the eigenvalues of $m$
  for g$-$modes can be quite close to $0$ especially for
  higher radial orders (i.e. the anti-Sturmian property).
As the specific entropy profile $g(x)$ tends to constant $1$ and
  $\gamma$ approaches $\Gamma=4/3$ simultaneously, the absolute
  value of buoyancy frequency squared $|{\cal N}^2|$ becomes
  smaller and smaller, squeezing all eigenvalues $m$ of
  g$-$modes to converge towards zero.

As shown in Section 4.1, we reveal a type of unstable perturbation
  solutions with eigenvalue $m=0$ ($n=1/3$) in that very limit of
  $g(x)=1$ and $\gamma=4/3$.
Growing with time $t$, these are purely vortical perturbations
  with $\mathbf{e_{\theta}}$ and $\mathbf{e_{\varphi}}$ components
  [see eq. (\ref{curl})]; meanwhile mass conservation (\ref{eq34})
  should be satisfied in the absence of mass density perturbation
  (see Appendix A for details).
In other words, the three components of vorticity perturbation
  $(\nabla\times{\mathbf v_1})$ all behave in the same temporal
  manner of unstable growth [see eqns (\ref{Rcurl})
  and (\ref{wrot})] with arbitrary dependence of
  $(x,\ \theta,\ \varphi)$.
Physically, they describe unstable vortical convections and
  circulations in a dynamically collapsing stellar core.
It would be more convenient and precise to regard
  them as separate vorticity modes of instability.

Complementarily, we may equally view such vortical perturbations
  as infinitely degenerate g$-$modes with eigenvalue $m=0$ ($n=1/3$).
%{\it
In reference to equation (\ref{eq32}), we can imagine the
  dimensionless specific entropy $g(x)$ to approach constant $1$
  and the perturbation polytropic exponent $\gamma$ to approach
  the background $\Gamma=4/3$;
given the fact that perturbation variable
  $w_r-\mbox{d}(xw_t)/\mbox{d}x$ (related to $\mathbf{e_{\theta}}$
  and $\mathbf{e_{\varphi}}$ components of vorticity perturbation)
  does not vanish for g$-$modes, it is then clear that all
  eigenvalues of $m$ concentrate towards $0$.
In that limiting regime, all g$-$modes would become completely
  degenerate.
%}
This also explains why vortical perturbation solutions obtained
  in Section 4.1 can have arbitrary vortical configurations as
  long as that the mass conservation (\ref{eq34}) is satisfied
  without inducing mass density perturbation.
%{\it This is just due the fact that these infinite degenerated
%g$-$modes can be recombined to form any pattern satisfying
%condition (\ref{eq34}). }
Therefore we may state that amplitude growths of g$-$modes
  in general are primarily governed by the angular momentum
  conservation (see Appendix D).
This implies that perturbation amplitudes of g$-$modes
  are roughly inversely proportional to the radius of
  the collapsing stellar core.
This conclusion should remain valid for a stellar core collapse
  with background polytropic exponent $\Gamma\neq 4/3$, for
  there should also exist amplification of perturbations
  resulting from the angular momentum conservation.
In general, growth rates of g$-$modes are also modified by the
  variable specific entropy evolution/distribution $g(x)$ as
  well as the corresponding $\lambda$ value and the difference
  between the perturbation polytropic exponent $\gamma$ and
  the background polytropic exponent $\Gamma=4/3$.
Sufficiently high-order g$^{+}-$modes and all g$^{-}-$modes are
  unstable, that is, they grow faster than the temporal factor
  $|t|^{-1/6}$ for stable acoustic p$-$modes and f$-$modes
  associated with a self-similar stellar core collapse
  (GW; Lou \& Cao 2008).

Such g$-$mode instabilities mainly governed by the angular
  momentum conservation might have a significant influence
  on the pre-SN core collapse stage and may affect the
  appearance of the PNS born during the stellar core
  collapse and the emergence of rebound shock as well
  as an SN explosion.
%{\bf\emph{
The effects of these unstable g$-$modes and vortical
  circulation modes during a stellar core collapse might have
  been omitted in previous multi-dimensional simulation works
  (e.g. Buras et al. 2006a; Dessart et al. 2006; Nordhaus
  et al. 2010b), because of their characteristic feature
  of weak or no density fluctuations.
Many numerical simulations show that the neutrino luminosity
  is not high enough to trigger an SN explosion.
It is considered physically plausible that the neutrino
  driven convection plays a key role in an SN explosion.
Our fast growing hydrodynamic convective g-modes and/or
  vorticity modes during a stellar core collapse, if
  properly included, might supply additional strengths
  for the PNS neutrino convection and circulation.
%    }}
These unstable perturbations will naturally destroy the
  spherical symmetry of the collapsing stellar core and
  give rise to subsequent aspherical nonlinear dynamic
  interactions.
%{\bf
The resulting asymmetric disturbances might actually be
  connected to and affect the initiation and evolution of
  SASI, although SASI is primarily of acoustic nature and
  operates between PNS and the stalled bounce shock.
In fact, g$-$mode oscillations of a PNS may also accompany
  SASI during the bounce phase (e.g. Burrows et al. 2006).
%  }
Our internal g$-$mode, vortical and convective instabilities
  exist during the core collapse irrespective of details of
  the specific entropy distribution/evolution.

Among all these unstable g$-$modes, those with $l=1$ allow
  a non-vanishing bulk flow velocity at the very centre of
  the progenitor stellar core.
These unstable g$-$modes may contribute directly or indirectly
  to the observed high peculiar speed of a new-born PNS
  (remnant of a stellar core collapse), which may reach
  values of several hundred to more than one thousand
  kilometers per second.
As an estimation, we assume the initial gas bulk flow velocity
  of the stellar core relative to the center of mass of
  the entire star to be $\sim 200\ {\rm m\ s}^{-1}$,
% on average,
  and the radius of the collapsing PNS contracts by a factor
  of $\sim 10^{-3}$, then the stellar core will gain a velocity
  around $\sim 200\ {\rm km\ s}^{-1}$ as a result of the
  rapid growth of $l=1$ g$-$mode perturbations, provided that the
  linear approximation still remains valid.\footnote{For a collapsed
  stellar core temperature of several times $10^{11}$K (e.g. Bethe
  1990), the PNS core sound speed would be in the order of
  $\sim 3\times 10^4\hbox{ km s}^{-1}$ and the estimate based
  on the linear approximation here should be justifiable. }
The smaller this contraction factor, the higher the PNS kick
  speed; the phase of PNS $l=1$ g$-$mode oscillation, the
  proper timing of stellar core collapse and the overall
  flow environment together should be also important to
  determine the net kick efficiency.
The initiation of such a collapsing PNS kick may actually
  occur prior to the emergence of a bounce shock.
%{\it
%{\bf
Recent numerical simulation studies suggest that the
 neutron star kick may be produced by the gravitational
 pull of the post-bounce anisotropic slow-moving ejecta
 (see figure 7 of Scheck et al. 2006) which may result
 from the neutrino-driven convection
%and a strong neutrino-driven wind
%that dissipates acoustic waves
 (e.g. Scheck et al. 2004; Nordhaus et al. 2010a).\footnote{
Parts of the work by Scheck et al. (2006) have already
  been presented in Scheck et al. (2004), but a detailed
  description of both their methods and results are given
  in the former. }
For example, based on extensive 2D model simulations with
 approximate neutrino transport and boundary conditions that
 parameterize the effects of the contracting NS and strong
 enough neutrino heating for neutrino-driven SN explosions,
 Scheck et al. (2006) compared their results with those of
 Burrows et al. (2006).
Started 20 ms after the core bounce, the simulation results of
 Scheck et al. (2006) can lead to a dominance of dipole ($l=1$)
 and quadruple ($l=2$) modes in the explosion ejecta, provided
 that the onset of the SN explosion is sufficiently slower
 than the growth time scale of the low-mode instability;
 they show that global anisotropies and large NS kicks can be
 obtained naturally in the framework of the neutrino-driven SN
 explosion mechanism due to the symmetry breaking by non-radial
 hydrodynamic instabilities, without resorting to rapid rotation,
 large pre-collapse perturbations in the iron core, strong
 magnetic fields, anisotropic neutrino emission associated
 with exotic neutrino properties, or jets etc.
As a complement to the work of Scheck et al. (2006),
 Nordhaus et al. (2010a) performed a 2D axisymmetric
 radiation-hydrodynamic simulation of the collapse of
 a non-rotating 15$M_{\odot}$ progenitor core and can
 naturally capture the PNS formation and subsequent
 off-center acceleration during a delayed, neutrino-driven,
 anisotropic SN explosion (their SN explosion is artificially
 induced by including additional neutrino luminosity to the
 calculation); at the end of their simulation, the PNS has
 achieved a velocity of $\sim 150\hbox{ km s}^{-1}$ and is
 still accelerating at $\sim 350\hbox{ km s}^{-2}$.
In the simulation of Burrows et al. (2006), the acoustic energy
 input from PNS g-mode oscillations was crucial for the explosion
 of an 11$M_{\odot}$ progenitor.
Nonspherical accretion was found to give rise to core g-mode
 oscillations at late times ($\gsim 300-500$ms) after core
 bounce, providing a significant amount of acoustic power for
 SN shock.
Actually, g-mode oscillations are also present in the outer
 layer of a NS according to Scheck et al. (2006), but the
 amplitudes are modest and do not lead to the strong effects
 of Burrows et al. (2006).
It is possible that Scheck et al. (2006) simulations might
 underestimate g-mode effects which would require the inclusion
 of the entire neutron star without excising the central core
 and thus the dynamic coupling among accretion, core motion,
 and deep g-mode generation in a self-consistent manner.
According to our estimates above, the core collapse hydrodynamic
 convective instabilities may also have a significant impact on
 the development of the anisotropic ejecta and the asymmetric
 bounce shock, and may also contribute directly to the PNS
 kick, providing a pre-existing background velocity during
 the formation of a nascent PNS.
It would be desirable that these sources of g$-$mode instabilities
 during core collapse be further explored by 3D numerical simulations.
% }{missed in the above numerical simulation works, in which
% no initial convection perturbation is explicitly put in.}
In addition, those unstable g$-$modes with $l=2$ might induce a
  collapsing stellar core to break up into two pieces, forming
  binary PNSs or other combinations around the centre of the
  pre-SN progenitor, while g$-$modes with higher $l$ and under
  favorable conditions may violently destroy the stellar core,
  shredding it into many smaller pieces and thus leading to the
  destruction of a PNS.\footnote{Although highly speculative,
  the absence of evidence over more than two decades for a
  central compact object in the remnant of SN1987A might be
  indicative of such a possibility of broken core.
  Admittedly, there could be other alternatives to resolve
  this issue. }
%{\bf
The possibility of such a situation might be low,
  as several multi-dimensional numerical simulations
  show that low$-l$ modes appear most likely dominant.
%   }

Based on our model analysis, the vorticity modes of
  instability appear unavoidable during a stellar core collapse;
  this would happen irrespective of the specific entropy
  evolution/distribution $g(x)$ including the conventional
  polytropic case of $g(x)=1$.
Depending on sources of seed fluctuations, rapid core spin,
  circulations of various scales, and vortical convections would
  go along with the stellar core collapse in a generic manner.
Physically, such 3D vortical motions can naturally lead to
  core convective turbulence and thus MHD dynamo actions for
  producing intense magnetic fields in PNSs and in stellar
  mass black holes.
Conceptually, we may specifically separate the stellar core
  rotation from the rest of radial vorticity perturbation
  [see eq. (\ref{wrot})] and envision the scenario of
  nonlinear Rossby waves (e.g. Haurwitz 1940; Lou 1987, 2001).

%\section{Conclusions and Discussion}
\section{Summary and Conclusions}

%In this paper, w
3D general polytropic
  perturbations in a $\Gamma=4/3$ self-similar collapsing
%relativistically hot
  stellar core have been examined.
% of a massive progenitor star.
We combine two recent generalized formulations
  adopted by Cao \& Lou (2009, 2010)
%\citet{b2} and \citet{b4}
  together to explore even more general cases
  of 3D perturbations.
The specific entropy evolution/distribution profile $g(x)$
  is allowed to vary from the stellar centre to its outer
  contracting boundary.
The special subcase of uniform $g(x)=1$ corresponds to a
  conventional polytropic gas sphere (GW; Lou \& Cao 2008).
The polytropic exponent $\gamma$ of 3D perturbations is
  generally allowed to be different from the background
  polytropic exponent $\Gamma=4/3$.
Not surprisingly, these general conditions give a non-zero
  Brunt-V$\ddot{\rm a}$is$\ddot{\rm a}$l$\ddot{\rm a}$ buoyancy
  frequency squared ${\cal N}^2$ as defined by equation (\ref{BV2}),
  and allow for two groups of g$-$mode perturbations to occur, viz.
  g$^{+}-$modes and g$^{-}-$modes under proper situations.
Meanwhile, acoustic p$-$modes and f$-$modes can also be clearly
  identified from the results of numerical calculations; again,
  they remain stable.

%{\it
We first revealed a class of unstable radial vorticity
  perturbation solutions that are incompressible and can be
  separated out independently from the rest of compressible
  perturbation variables under fairly general situations,
  i.e. irrespective of the specific entropy
  evolution/distribution profile $g(x)$.
Such unstable vortical perturbations include the spin of the
  collapsed stellar core and all kinds of sufficiently slow vortex
  motions tangent to spherical layers in the stellar core.
Their velocity amplitudes are inversely proportional to
  the radius of the collapsed stellar core, as a result
  of the angular momentum conservation.
This appears to be a natural mechanism for producing
  fast spins of PNSs or stellar mass
  black holes in the eventual nonlinear evolution.
Conceptually, one may further separate the core spin and
  other radial vorticity perturbations; their mutual
  nonlinear interactions and evolution may be perceived.
We therefore suggest that Rossby waves may occur
  and be amplified in nonlinear dynamic processes.
Such Rossby waves may be also connected with the
  spin down process of a nascent neutron star.
If the collapsing stellar core is magnetized, significant
  MHD dynamo actions may also take place to amplify
  magnetic fields during the core collapse process involving
  considerable turbulent convections and circulations.
%}

By the inverse iteration method (e.g. Wilkinson 1965), we then
  solve the remaining general polytropic compressible perturbation
  equations (\ref{eq30})$-$(\ref{eq32}) numerically with specified
  boundary conditions (\ref{BCs}), and examine the eigenvalues and
  eigenfunctions of perturbations, including acoustic p$-$modes,
  surface f$-$modes and internal g$-$modes, which can be further
  divided into two subclasses: internal g$^{+}-$modes and internal
  g$^{-}-$modes.
Sufficiently high radial order g$^{+}-$modes and all g$^{-}-$modes
  show instabilities during the dynamic core collapse process,
  leading to convective motions of circulation within the
  collapsing stellar core.
In contrast, acoustic p$-$modes, f$-$modes and sufficiently
  low order g$^{+}-$modes remain stable.
For a certain specific entropy evolution/distribution,
  there may exist g$-$modes that are drastically oscillatory
  deeply trapped in the collapsing stellar core.

The mechanism of unstable growths of g$-$modes is identified.
 We realize that the fast temporal increase in amplitudes
 of g$-$modes are mainly governed by the angular momentum
 conservation.
%{\it
It is easy to understand this intuitively, for g$-$mode
  perturbations are mainly convective motions inside
  the dynamically collapsing stellar core, in which
  the gas runs in circulations.
%When the circle of flow contracts,
During the core collapse, the velocity of
  the gas flow will increase accordingly
%{\bf
  (see Appendix D for more details).
%}
%{\bf
In a more realistic model for stellar core collapse,
  both $\kappa$ and $\Gamma$ may vary in time.
However, during a long period before the end of the collapse of
  stellar core, numerical simulations (e.g. Burrows \& Lattimer
  1986; Van Riper \& Lattimer 1981; Bethe 1990) show that the
  variations of $\kappa$ and $\Gamma$ are slow and not obvious,
  and our conclusions may still be valid approximately.
  %}
The amplitude of g$-$mode perturbations are more or less inversely
  proportional to the radius of collapsing stellar core.
When the specific entropy distributes uniformly and the
  perturbation polytropic exponent $\gamma=\Gamma=4/3$, this
  temporal increasing factor of perturbation amplitude
  becomes exact, and all the g$-$modes become completely
  degenerate in this regard.
Alternatively, we can regard these unstable modes as 3D vorticity
  perturbations constrained by the mass conservation.
They may bear physical consequences for the central compact
  object, be it PNSs or stellar mass black holes.

Given the idealizations and approximations of our model,
  these unstable g$-$modes may offer certain valuable
  clues for SN model simulations.
They affect the geometry of the rebound shock wave
  emerged after the core bounce because of asymmetries
%{\bf\emph{
  and enhance the neutrino driven PNS convection.
Most simulation works do not explicitly
  include these convective g-mode perturbations (e.g.
  Dessart et al. 2006; Nordhaus et al. 2010b).
Buras et al. (2006b) added in their simulation an arbitrary
  density perturbation before the core collapse, and do not
  observe significant difference.
The density perturbations, however, are likely more
  acoustic and grow slowly, because the convective
  g-modes involve weak density fluctuations.
Among the various g-modes,
%  }}
  the $l=1$ g$-$modes with central
  movement may possibly be responsible for the origin of the
  high kick speed that a new-born neutron star acquires,
%  {\bf
  or serve as the source of producing anisotropic
  stellar ejecta which may possibly gravitationally
  accelerate a nascent PNS (Scheck et al. 2006;
  Nordhaus et al. 2010a).
%   }. {\it
Nonlinear evolution of those unstable g$-$modes with
  $l$ higher than $1$ may possibly break a collapsing
  core into two or multiple pieces, forming a binary
  system of neutron stars or more fragments as to even
  ultimately prevent the formation of a neutron star.
%}

Instabilities of radial component vorticity perturbation can lead
  to rapid rotations of central compact objects, Rossby waves,
  and circulations over spherical surfaces inside a core.
Instabilities of 3D vorticity perturbation when $g(x)=1$ and
  $\gamma=\Gamma=4/3$ can give rise to convective turbulence
  in the core and MHD dynamo actions to sustain violent
  magnetic activities.

\section*{Acknowledgments}

This research was supported in part by Tsinghua Centre for
 Astrophysics, by the National Natural Science Foundation
 of China grants 10373009, 10533020, 11073014 and J0630317
 at Tsinghua University, by MOST grant 2012CB821800,
%  射电波段的前沿天体物理课题及FAST早期科学研究 公示内容
 by Tsinghua University Initiative Scientific Research Program,
 and by the Yangtze Endowment, the SRFDP 20050003088 and
 200800030071, and the Special Endowment for Tsinghua College
 Talent (Tsinghua XueTang) Program from the Ministry of
 Education at Tsinghua University.

%\newpage
\appendix

%\section{Necessity of Equation (35)\\
%\qquad\qquad\qquad\quad for $\gamma=4/3$, $g(x)=1$ and $m=0$}
\section{Equation (35) is necessary given $\gamma=4/3$, $g(x)=1$ and $m=0$}

%{\bf Not enough number of equations? Not a comfortable analysis!}

In the special case of $\gamma=4/3$, $g(x)=1$ and $m=0$ for a
  conventional polytropic gas, equation (\ref{eq32}) is
  automatically satisfied. In the first subcase of $n=0$, equation
  (\ref{eq34}) comes out automatically from equation (\ref{eq30}).
When we focus on the other subcase of $n=1/3$ satisfying the same
  requirement $m=0$, we derive from equation (\ref{eq31}) the
  following relation
\begin{equation}\label{eqA1}
\psi_1=-2\bigg[\frac{f}{x^2}\frac{\mbox{d}}{\mbox{d}x}(x^2w_{r})
-\frac{l(l+1)f}{x}w_{t}+3f'w_{r}\bigg]\ ;
\end{equation}
meanwhile equation (\ref{eq30}) yields
\begin{equation}\label{eqA2}
\frac{\mbox{d}^2(x\psi_1)}{\mbox{d}x^2}
=\bigg[\frac{l(l+1)}{x^2}-3f^2\bigg](x\psi_1)\ .
\end{equation}
%{\it
In the interval $(x_b,\ +\infty)$, we define $f(x)=0$. To
  avoid singularity, we naturally require $x\psi_1$ to
  remain finite at $x=0$ and $x\rightarrow +\infty$.
Equation (\ref{eq30}) should be regarded as a
  Sturm-Liouville problem within the interval
  $[0,\ +\infty)$ for determining the
  eigenvalues $l(l+1)$.
%, or, equivalently, $l$. {\bf Not really!}
The eigenvalues $l$ are determined by the specific profile
  of the dynamic background density function $f(x)$ and in
  general cannot be exactly equal to an integer.
This means that in general we could only have the trivial
  solution $x\psi_1=0$ for integral values of $l$.
Another argument proving the necessity of $x\psi_1=0$ can
  be made in the following for those cases $l\ge 2$.
%}
Since $x\psi_1$ should remain finite at $x=0$ to avoid singularity
  of a proper perturbation solution, $x\psi_1\propto x^{l+1}$ should
  hold as $x\rightarrow 0^+$.
We also need to require $x\psi_1\propto x^{-l}$ when $x\rightarrow
  +\infty$ so that $\psi_1$ approaches zero at infinity.
This means that the algebraic sign of
  $\mbox{d}(x\psi_1)/\mbox{d}x$ should necessarily
  reverse when $x$ varies from $0^+$ to $+\infty$.
Suppose that $x\psi_1$ is positive when $x\rightarrow 0^+$,
  then its first derivative is also positive.
However, from the numerical results of $f(x)$ it is found
  that for $l\ge 2$, inequality $l(l+1)/x^2-3f^2>0$ holds
  in the entire interval $(0,+\infty)$, which indicates
  that $\mbox{d}^2(x\psi_1)/\mbox{d}x^2$ remains positive
  as long as $x\psi_1$ is positive.
There seems no chance for the sign of
  $\mbox{d}(x\psi_1)/\mbox{d}x$ to reverse.
%{\bf This may not be a sure argument!}
Then $x\psi_1$ must vanish. In conclusion, we show that
  $x\psi_1=0$, and according to equation (A1), the
  necessity of equation (\ref{eq34}) is warranted.

%\section{Orthogonality of eigenfunctions with different eigenvalues $m$}
\section{Orthogonal set of eigenfunctions with distinct eigenvalues $m$}

Perturbation eigenfunctions with different eigenvalues of
  $m=3\lambda n(3n-1)/2$ can be proven to be mutually
  orthogonal to each other.
In the proof below, we use superscripts $^{(1)}$ and $^{(2)}$
  to distinguish different eigenfunctions and eigenvalues.
From perturbation equations (\ref{eq22}) and (\ref{eq23}), we
  readily find
\begin{equation}\label{eqB1}
\nabla^2\frac{\partial\psi_1}{\partial\tau}
=2\nabla\cdot(f^3\mathbf{v_1})\ ;
\end{equation}
a substitution of equations (\ref{eq23}) and (\ref{eq24})
  into equation (\ref{eq21}) then yields
\[\frac{4}{3}mf^3\mathbf{v_1}
=\nabla\Big[\gamma gf^4(\nabla\cdot\mathbf{v_1})
+\mathbf{v_1}\cdot\nabla(gf^4)\Big]\]
\begin{equation}\label{eqB2}
\qquad\ \quad-\Big(\nabla\cdot\mathbf{v_1}
+\mathbf{v_1}\cdot\frac{3\nabla f}{f}\Big)\nabla(gf^4)
+2f^3\nabla\frac{\partial\psi_1}{\partial\tau}\ .
\end{equation}
We now evaluate the spatial integral below with the help
  of equations (\ref{eqB1}) and (\ref{eqB2}), namely
\[\quad\frac{4}{3}m^{(1)}\int f^3\mathbf{v_1}^{(1)}
\cdot\mathbf{v_1}^{(2)}
\mbox{d}V\]
\[=-\int\Big[\gamma gf^4\nabla\cdot\mathbf{v_1}^{(1)}
+\mathbf{v_1}^{(1)}\cdot\nabla(gf^4)
\Big]\nabla\cdot\mathbf{v_1}^{(2)}\mbox{d}V\]
\[\quad -\int\bigg[\nabla\cdot\mathbf{v_1}^{(1)}
+\mathbf{v_1}^{(1)}\cdot\frac{3\nabla f}{f}
\bigg]\nabla(gf^4)\cdot\mathbf{v_1}^{(2)}\mbox{d}V\]
\[\quad -\int2\frac{\partial\psi_1^{(1)}}{\partial\tau}
\nabla\cdot\big[f^3\mathbf{v_1}^{(2)}\big]\mbox{d}V\]
\[=-\int \bigg\lbrace\gamma gf^4\big[\nabla\cdot\mathbf{v_1}^{(1)}\big]
(\nabla\cdot\mathbf{v_1}^{(2)})\]
\[\qquad+\big[\mathbf{v_1}^{(1)}\cdot\nabla(gf^4)\big]
(\nabla\cdot\mathbf{v_1}^{(2)})\]
\[\qquad +\big[\mathbf{v_1}^{(2)}\cdot\nabla(gf^4)\big]
(\nabla\cdot\mathbf{v_1}^{(1)})\]
\[\qquad+\big[\mathbf{e_r}\cdot\mathbf{v_1}^{(1)}\big]
\big[\mathbf{e_r}\cdot\mathbf{v_1}^{(2)}\big]\frac{3\nabla
f\cdot\nabla(gf^4)}{f}\]
\begin{equation}\label{eqB3}
\qquad-\Big(\nabla\frac{\partial\psi_1^{(1)}}{\partial\tau}
\Big)\cdot\Big(\nabla\frac{\partial\psi_1^{(2)}}
{\partial\tau}\Big)\bigg\rbrace\mbox{d}V\ ,
\end{equation}
where $dV$ is the volume element for the spatial integration.
Noting that the RHS of equation (\ref{eqB3}) is symmetric with
 respect to two superscripts $^{(1)}$ and $^{(2)}$, we have
\begin{equation}\label{eqB4}
\big[m^{(1)}-m^{(2)}\big]\int f^3\mathbf{v_1}^{(1)}
\cdot\mathbf{v_1}^{(2)}\mbox{d}V=0\ ,
\end{equation}
and the orthogonality of eigenfunctions with different
  eigenvalues of $m$ is therefore proven.

\section{Inverse Iteration Scheme for
  Solving Linear ODEs (30)$-$(32)}

We give here a procedure description of the inverse iteration
  scheme (e.g. Wilkinson 1965) as adapted to our numerical
  computations of 3D general polytropic perturbations in a
  self-similar dynamic stellar core collapse of spherical
  symmetry and $\Gamma=4/3$.

First, we specify a chosen form of the specific entropy function
  $g(x)$ and employ the fourth-order Runge$-$Kutta scheme (e.g.
  Press et al. 1996) to integrate background ODE (\ref{GLane})
  for $f(x)$ from $x=0^+$ and thus obtain the dynamic background
  function $f(x)$ with the initial or boundary conditions
  $f(0)=1$ and $f'(0)=0$;
  this $f(x)$ is closely related to the background mass density
  profile.
For $\Gamma=4/3$, the contracting boundary of the stellar core
  $x_b$ is determined by the realizable requirement $f(x_b)=0$
  in a parameter range of $0<\lambda\leq\lambda_c$ for a proper
  $\lambda$ value.
Our goal is to reliably solve linear ODEs
  (\ref{eq30})$-$(\ref{eq32}) for 3D general polytropic compressible
  perturbations within the interval $[0,\ x_b]$ by determining
  the eigenvalues of $m=3\lambda n(3n-1)/2$ and the corresponding
  perturbation eigenfunctions for given $g(x)$, $f(x)$, $\lambda$,
  $x_b$, $l$ and $\gamma\neq\Gamma=4/3$ in general.

In order to implement the inverse iteration scheme, we need
  to discretize linear ODEs (\ref{eq30})$-$(\ref{eq32})
  with a proper mesh in the interval $(0,\ x_b)$.
More specifically, we first divide the entire interval
%{\bf averagely?}
  evenly into $N$ small intervals and define the interval
  size $\Delta x\equiv h=x_b/N$; we can then construct a
  vector $\mathbf{w}$ of length $2(N-1)$ for $w_r$ and
  $w_t$ arranged together contiguously in order, namely
\[\mathbf{w}=\Big[w_r(h),w_r(2h),\cdots w_r((N-1)h),\]
\begin{equation}
\qquad\qquad w_t(h),w_t(2h),\cdots w_t((N-1)h)\Big]^T\ .
\end{equation}

We also need to cast every differential
  operator into a matrix form.
The matrix for a function serving as coefficients in
  the ODEs, say $y(x)$, would then take the form of
\begin{equation}
Y=Diag\Big[y(h),y(2h),\cdots y((N-1)h)\Big]\ ,
\end{equation}
which is a matrix of dimension $(N-1)$. By notation `$Diag$', we
  refer to a diagonal matrix. The matrices $D$ and $D^2$ for the
  differential operators $\mbox{d}/\mbox{d}x$ and
  $\mbox{d}^2/\mbox{d}x^2$ are respectively
\begin{equation}
D=\frac{1}{h}\left(\begin{array}{ccccc}
 1&  &     &     &     \\
-1& 1&     &     &     \\
  &-1&  1  &     &     \\
  &  &\cdot&\cdot&\cdot\\
  &  &     &  -1 &  1  \\
\end{array}\right)\
\end{equation}
and
\begin{equation}
D^2=\frac{1}{h^2}\left(\begin{array}{ccccc}
-2& 1&      &     &     \\
 1&-2&   1  &     &     \\
  & 1&  -2  &  1  &     \\
  &  & \cdot&\cdot&\cdot\\
  &  &      &  1  &  -2 \\
\end{array}\right)\ ,
\end{equation}
both are of dimension $(N-1)$.
  For different boundary conditions, these matrices
  above may need modifications accordingly.
%{\bf Curious about this. At $0$ and at $Nh$?}
Note that ODE (\ref{eq31}) can be substituted into ODE
  (\ref{eq30}) to eliminate the dependent variable $n\psi_1$.
With these procedures, we could finally achieve the discretization
  scheme of linear ODEs (\ref{eq30})$-$(\ref{eq32}).
The resulting finite difference scheme can
  be simply cast into the following form of
\begin{equation}\label{C5}
\left(\begin{array}{cc}A_1&A_2\\A_3&A_4\end{array}\right)\mathbf{w}
=A\mathbf{w}=m\mathbf{w}\ ,
\end{equation}
in which $A_i\ (i=1,2,3,4)$ are four matrices all of dimension
  $(N-1)$ that do not involve the eigenvalue parameter $m$;
  here, eigenvalue parameter $m$ appears only on the RHS and
  is to be determined for discrete eigenvalues.

%{\bf Would it be possible to provide the
%specific forms of these for matices?}

We can then apply the inverse iteration method to solve this
  eigenvalue problem.
We begin to specify an arbitrary vector $\mathbf{w}_0$ and guess
  an approximate eigenvalue $m_0$, and then rewrite the matrix
  equation (\ref{C5}) as
\begin{equation}
(A-m_0)\mathbf{w}=(m-m_0)\mathbf{w}\ .
\end{equation}
Suppose that all the eigenvalues and eigenfunctions of matrix
  equation (\ref{C5}) are respectively $m_i$ and $\mathbf{w}_i\
  (i=1,2,\cdots)$, among which the value of $m_j$ is closest to the
  initially chosen value $m_0$.
Then, $\mathbf{w}_0$ can be expanded as $\mathbf{w}_0=\sum_i
  a_i\mathbf{w}_i$, where $a_i$ stands for a set of coefficients,
  and it is thus easy to demonstrate
\begin{equation}
(A-m_0)^{-n}\mathbf{w_0}\rightarrow a_j(m_j-m_0)^{-n}\mathbf{w_j}
\quad\mbox{as}\quad n\rightarrow\infty\ .
\end{equation}
%{\bf Please clarify.}
Therefore, by iterating an enough number of times, we can select
  the eigenfunction $\mathbf{w}_j$ and determine the corresponding
  eigenvalue $m_j=(m_j-m_0)+m_0$.
Changing the value of $m_0$, we would be able to select other
  eigenfunctions and eigenvalues of the perturbation problem.

%\section{Connection between angular momentum conservation
%and amplitude increase of g-modes during a core collapse}
\section{Magnitude growth of g-modes during a
 core collapse by angular momentum conservation}

The angular momentum in a certain spatial volume
  $V$ of first-order precision can be cast in the
  integral form of
%{\bf The sign of the cross product; email sent to Lian Biao.}
\[\mathbf{J}_V=\int_V\rho(\mathbf{r}\times\mathbf{u})
d^3\mathbf{r}=\Big(\frac{\kappa_c}{\pi G}\Big)^{3/2}
\dot{a}a\int_Vf^3(\mathbf{x}\times\mathbf{v_1})d^3\mathbf{x}\]
\begin{equation}
\qquad\qquad ={\cal A}e^{\tau/3}\int_Vf^3(\mathbf{x}
  \times\mathbf{v_1})d^3\mathbf{x}\ ,
\end{equation}
where ${\cal A}$ is a constant. Taking the derivative of
  the angular momentum in terms of the logarithmic time
  $\tau=-\ln|t|$ and using equation (\ref{eq21}), we obtain
\[\frac{d\mathbf{J_V}}{d\tau}={\cal A}e^{\tau/3}
 \int_Vf^3\bigg[\mathbf{x}\times\bigg(\frac{\partial}{\partial\tau}
 -\frac{1}{3}\bigg)\mathbf{v_1}\bigg]d^3\mathbf{x}\]
\begin{equation}
={\cal A}e^{\tau/3}\int_V\bigg[\mathbf{x}\times\bigg(\frac{gf^4}
 {4\lambda}\nabla\beta_1+\frac{f^3}{3\lambda}
 \nabla\psi_1\bigg)\bigg]d^3\mathbf{x}\ .
\end{equation}
From equations (\ref{eq22})$-$(\ref{eq24}),
  it immediately follows that
\begin{equation}
\nabla^2\psi_1=2\int\nabla\cdot(f^3\mathbf{v_1})d\tau\ ,
\end{equation}
\begin{equation}
\beta_1=\frac{2}{3}\int\Big[\gamma\nabla\cdot
(f^3\mathbf{v_1})+\mathbf{v_1}\cdot \Big((4-3\gamma)\frac{\nabla
f}{f} +\frac{\nabla g}{g}\Big)\Big]d\tau.
\end{equation}
Since $\nabla\psi_1$ is a potential vector field,
  integral (D3) can be further simplified as
\begin{equation}
\nabla\psi_1=2\int(f^3\mathbf{v_1})_pd\tau\ ,
\end{equation}
where $(f^3\mathbf{v_1})_p$ stands for the potential
  flow part of the vector field $f^3\mathbf{v_1}$.
Since for g$-$modes the vector field $f^3\mathbf{v_1}$ (which is
  proportional to the momentum density) is close to divergence free
  [which becomes exact when $g(x)=1$ and $\gamma=4/3$, see equation
  (\ref{eq34})], $\nabla\psi_1$ of g-modes should be quite small.
For nearly constant $g(x)$ and $\gamma\approx 4/3$, it is
  clear that $\beta_1$ is also a small quantity.
We therefore have an approximate angular momentum conservation law
\begin{equation}
\frac{d\mathbf{J_V}}{d\tau}\approx0\ ,
\end{equation}
which indicates
  $(\partial/\partial\tau-1/3)\mathbf{v_1}\approx0$.
It is thus the conservation of angular momentum that
  mainly governs the temporal increase of the
  velocity perturbation amplitude.

\bsp
\label{lastpage}

\begin{thebibliography}{}

\bibitem[\protect\citeauthoryear{Andersson}{1999}]{AnderssonA}
Andersson N., Kokkotas K. D., Stergioulas N., 1999a, ApJ, 516, 307
%-314; Andersson, Nils; Kokkotas, Kostas D.; Stergioulas, Nikolaos
%On the Relevance of the R-Mode Instability
%for Accreting Neutron Stars and White Dwarfs

\bibitem[\protect\citeauthoryear{Andersson}{1999}]{AnderssonB}
Andersson N., Kokkotas K. D., Schutz B. F., 1999b, ApJ, 510, 846
%--853; Gravitational Radiation Limit on the Spin of Young
%Neutron Stars
%A newly discovered instability in rotating neutron stars, driven
%by gravitational radiation reaction acting on the stars' r-modes,
%is shown here to set an upper limit on the spin rate of young
%neutron stars. We calculate the timescales for the growth of
%linear perturbations due to gravitational radiation reaction, and
%for dissipation by shear and bulk viscosity, working to second
%order in a slow-rotation expansion within a Newtonian polytropic
%stellar model. The results are very temperature-sensitive: in hot
%neutron stars (T>10^9 K), the lowest-order r-modes are unstable,
%while in colder stars they are damped by viscosity. These
%calculations have a number of interesting astrophysical
%implications. First, the r-mode instability will spin down a newly
%born neutron star to a period close to the initial period inferred
%for the Crab pulsar, probably between 10 and 20 ms. Second, as an
%initially rapidly rotating star spins down, an energy equivalent
%to roughly 1 of a solar mass is radiated as gravitational waves,
%which makes the process an interesting source for detectable
%gravitational waves. Third, the r-mode instability rules out the
%scenario in which millisecond pulsars are formed by
%accretion-induced collapse of a white dwarf; the new star would be
%hot enough to spin down to much slower rates. Stars with periods
%less than perhaps 10 ms must have been formed by spin-up through
%accretion in binary systems, where they remain colder than the
%Eddington temperature of about 10^8 K. More accurate calculations
%will be required to define the limiting spin period more reliably,
%and we discuss the importance of the major uncertainties in the
%stellar models, in the initial conditions after collapse, and in
%the physics of cooling, superfluidity, and the equation of state.
%

\bibitem[\protect\citeauthoryear{Atoyan}{1999}]{atoyan}Atoyan
B. M., 1999, A\&A, 346, L49
%regarding the inference of the initial
%spin period of the Crab pulsar

\bibitem[\protect\citeauthoryear{Bethe}{1990}]{bethe} Bethe
H. A., 1990, Rev. Mod. Phys., 62, 801

\bibitem[\protect\citeauthoryear{Bethe et al.}{1979}]{b17} Bethe
H. A., Brown G. E., Applegate J., Lattimer J. M., 1979, Nucl.
Phys. A, 324, 487

\bibitem[\protect\citeauthoryear{Blondin, Mezzacappa
\& DeMarino}{2003}]{b10}
Blondin J. M., Mezzacappa A., DeMarino C., 2003, ApJ, 584, 971

\bibitem[\protect\citeauthoryear{Blondin \& Mezzacappa}{2006}]{Blondin-ApJ}
 Blondin J. M., Mezzacappa A., 2006, ApJ, 642, 401

\bibitem[\protect\citeauthoryear{Blondin \& Mezzacappa}{2007}]{Blondin-Nature}
 Blondin J. M., Mezzacappa A., 2007, Nature, 445, 58

\bibitem[\protect\citeauthoryear{Blondin \& Mezzacappa}{2007}]{BlondinShaw-ApJ}
 Blondin J. M., Shaw S., 2007, ApJ, 656, 366

\bibitem[\protect\citeauthoryear{Bruenn}{1985}]{b18} Bruenn S. W.,
1985, ApJ, 58, 771

\bibitem[\protect\citeauthoryear{Bruenn}{1989a,b}]{b14}Bruenn S. W.,
1989a, ApJ, 340, 955

\bibitem[\protect\citeauthoryear{Bruenn}{1989a,b}]{b15}Bruenn S. W.,
1989b, ApJ, 341, 385

\bibitem[\protect\citeauthoryear{Buras et al.}{2006a}]{Buras2006a}Buras R.,
 Rampp M., Janka H. Th., Kifonidis K., 2006a, A\&A, 447, 1049

\bibitem[\protect\citeauthoryear{Buras et al.}{2006b}]{Buras2006b}Buras R.,
 Janka H. Th., Rampp M., Kifonidis K., 2006b, A\&A, 457, 281

\bibitem[\protect\citeauthoryear{Burrows \& Hayes}{1996}]{BH1996}
Burrows A., Hayes J., 1996, Phys. Rev. Lett., 76, 352

\bibitem[\protect\citeauthoryear{Burrows \& Lattimer}{1983}]{b04}
Burrows A., Lattimer J. M., 1983, ApJ, 270, 735

\bibitem[\protect\citeauthoryear{Burrows \& Lattimer}{1986}]{b01}
Burrows A., Lattimer J. M., 1986, ApJ, 307, 178

\bibitem[\protect\citeauthoryear{Burrows et al.}{2006}]{b12}Burrows A.,
Livne E., Dessart L., Ott C. D., Murphy J., 2006, ApJ, 640, 878

\bibitem[\protect\citeauthoryear{Burrows et al.}{2006}]{b13}Burrows A.,
Dessart L., Ott C. D., Livne E., 2007, Phys. Rep., 442, 23 Title:
% Multidimensional Radiation/Hydrodynamic Simulations of Proto-Neutron
%Star Convection
%Affiliation:
% AA(Department of Astronomy and Steward Observatory,
%University of Arizona, Tucson, AZ 85721 luc@as.arizona.edu
%burrows@as.arizona.edu), AB(Department of Astronomy and
%Steward Observatory, University of Arizona, Tucson, AZ 85721
%luc@as.arizona.edu burrows@as.arizona.edu),
%AC(Racah Institute of Physics, Hebrew University, Jerusalem,
%Israel eli@frodo.fiz.huji.ac.il), AD(Max-Planck-Institut
%für Gravitationsphysik, Albert-Einstein-Institut,
%Golm/Potsdam, Germany cott@aei.mpg.de)
%
%Based on multidimensional, multigroup, flux-limited-diffusion
%hydrodynamic simulations of core-collapse supernovae with the
%VULCAN/2D code, we study the physical conditions within and in
%the vicinity of the nascent proto-neutron star (PNS). Our
%numerical study follows the evolution of the collapsing envelope
%of the 11 Msolar model of Woosley & Weaver from ~200 ms before
%bounce to ~300 ms after bounce on a spatial grid that switches
%from Cartesian at the PNS center to spherical above a 10 km
%radius. As has been shown previously, we do not see any
%large-scale overturn of the inner PNS material. Convection,
%directly connected to the PNS, is found to occur in two distinct
%regions, between 10 and 20 km, coincident with the region of
%negative lepton gradient, and exterior to the PNS, above 50 km.
%Separating these two regions, an interface with no sizable
%inward or outward motion is the site of gravity waves,
%emerging at 200-300 ms after core bounce, excited by the
%convection in the outer convective zone. In the PNS
%convection is always confined within the neutrinospheric
%radii for all neutrino energies above just a few MeV. We
%find that such convective motions do not appreciably
%enhance the νe neutrino luminosity, and that they enhance
%the νˉe and ``νμ'' luminosities modestly by ~15% and
%~30%, respectively, during the first postbounce 100-200 ms.
%Moreover, we see no evidence of doubly diffusive
%instabilities in the PNS, expected to operate on diffusion
%timescales of at least a second, much longer than the
%millisecond timescale associated with PNS convection.
%PNS convection is thus found to be a secondary feature
%of the core-collapse phenomenon, rather than a decisive
%ingredient for a successful explosion.
%

\bibitem[\protect\citeauthoryear{Cao \& Lou}{2009}]{b2}Cao Y.,
 Lou Y.-Q., 2009, MNRAS, 400, 2032
%(Perturbation Analysis of a General Polytropic
%Homologously Collapsing Stellar Core)

\bibitem[\protect\citeauthoryear{Cao \& Lou}{2010}]{b4} Cao Y., Lou Y.-Q.,
2010, MNRAS, 403, 491
%(Adiabatic perturbations in homologous conventional
%polytropic core collapses of a spherical star)

\bibitem{bb45} Chandrasekhar S., 1939, An Introduction to the
Study of Stellar Structure, Dover Publications, Inc., London

\bibitem{bb46} Chandrasekhar S., 1961, Hydrodynamic and
Hydromagnetic Stability, Dover Publications, Inc., New York

\bibitem[\protect\citeauthoryear{Cowling}{1941}]{b5}
Cowling T. G., 1941, MNRAS, 101, 367
%(The non-radial oscillations of polytropic stars)

\bibitem[\protect\citeauthoryear{Cox}{1976}]{Cox76}
Cox J. P., 1976, ARA\&A, 14, 247
%Nonradial oscillations of stars: theories and observations

\bibitem[\protect\citeauthoryear{Dessart et al.}{2006}]{Dessart2006}
Dessart L., Burrows A., Livne E., Ott C.D., 2006, ApJ, 645, 534

\bibitem[\protect\citeauthoryear{Dimmelmeier et al.}{2008}]{b06}
Dimmelmeier H., Ott C. D., Marek A., Janka H. T., 2008,
Phys. Rev. D, 78, 064056
%The gravitational wave burst signal
%from core collapse of rotating stars

\bibitem{Dynamo1} Duncan R. C., Thompson C., 1992, ApJ, 392, L9
%Formation of Very Strong Magnetized Neutron Stars:
%Implications for Gamma Ray Bursts

\bibitem{b45} Eddington A. S., 1926, The Internal Constitution
 of the Stars, Cambridge University Press, Cambridge

\bibitem[\protect\citeauthoryear{Fern\`andez}{2010}]{fer}
Fern\`andez R., 2010, ApJ, 725, 1563
%regarding SASI spiral mode....

\bibitem[\protect\citeauthoryear{Foglizzo}{2001}]{f1}
Foglizzo T., 2001, A\&A, 368, 311

\bibitem[\protect\citeauthoryear{Goldreich \& Weber}{1980}]{b1}
Goldreich P., Weber S. V., 1980, ApJ, 238, 991
%(Homologously collapse stellar cores)

\bibitem[\protect\citeauthoryear{Goldreich, Lai \& Sahrling}{1996}]{b7}
Goldreich P., Lai D., Sahrling M., 1996, in Bahcall J. N., Ostriker J. P.,
eds, Unsolved Problems in Astrophysics. Princeton Univ. Press, Princeton

\bibitem[\protect\citeauthoryear{Haurwitz}{1940}]{Haurwitz40}
Haurwitz B., 1940, J. Marine Res., 3, 254
%related to global Rossby waves....

\bibitem[\protect\citeauthoryear{Hix et al.}{2003}]{b05} Hix W. R.,
Messer O. E. B., Mezzacappa A., Liebendorfer M., Sampaio J.,
Langanke K., Dean D. J., Martinez-Pinedo G., 2003, Phys. Rev.
Lett. 91, 201102
%Consequences of Nuclear Electron Capture in Core Collapse Supernovae
%

\bibitem[\protect\citeauthoryear{Hu \& Lou}{2009}]{HuLou09}
Hu R. Y., Lou Y.-Q., 2009, MNRAS, 396, 878
%Issue 2, 878-886; (2009arXiv:0902.3111)
%Magnetized Massive Stars as Magnetar Progenitors

\bibitem[\protect\citeauthoryear{Janka \& M\"uller}{1994}]{JM1994}
Janka H.-T., M\"uller E., 1994, A\&A, 290, 496

\bibitem[\protect\citeauthoryear{Iwakami et al.}{2008}]{Iwakami2008} Iwakami W.,
Kotake K., Ohnishi N., Yamada S., Sawada K., 2008, ApJ, 678, 1207

\bibitem[\protect\citeauthoryear{Lai}{2000}]{b8}Lai D.,
2000, ApJ, 540, 946

\bibitem[\protect\citeauthoryear{Lai, Chernoff \& Cordes}{2001}]{Laietal}
 Lai D., Chernoff D. F., Cordes J. M., 2001, ApJ, 549, 1111

\bibitem[\protect\citeauthoryear{Lai \& Goldreich}{2000}]{b9} Lai D.,
Goldreich P., 2000, ApJ, 535, 402

\bibitem[\protect\citeauthoryear{Lou}{1987}]{mRossby} Lou Y.-Q.,
1987, ApJ, 322, 862
%--869; Nonlinear Magnetohydrodynamic Waves i a Steady
%Zonal Circulation for a Shallow Fluid Shell on the
%surface of a Rotating Sphere
%

\bibitem[\protect\citeauthoryear{Lou}{1990}]{b21} Lou Y.-Q.,
1990, ApJ, 361, 527

\bibitem[\protect\citeauthoryear{Lou}{1991}]{b22} Lou Y.-Q.,
1991, ApJ, 367, 367

\bibitem{Lou1995a} Lou Y.-Q., 1995a, ApJ, 442, 401
%-404;  ``Gravity Waves in the Lower Solar Corona"
%

\bibitem{Lou1995b} Lou Y.-Q., 1995b, MNRAS, 276, 769
%``Gravito-Acoustic Wave Transformation in Stellar Atmospheres",
%Lou, Y.-Q., {\it Monthly Notices of the Royal Astronomical
%Society}, {\bf 276}, 769-784, 1995.
%oddjob machine directory agtm
%

\bibitem{Lou1996} Lou Y.-Q., 1996, Science, 272, 521
%``Trapped Coronal Magneto-Gravity Modes", Lou, Y.-Q.,
%{\it Science}, {\bf 272}, 521-523, 1996.
%

\bibitem{LouRossby} Lou Y.-Q., 2000, ApJ, 540, 1102
%ROSSBY-TYPE WAVE-INDUCED PERIODICITIES IN FLARE ACTIVITIES
%AND SUNSPOT AREAS OR GROUPS DURING SOLAR MAXIMA

\bibitem{LouTide} Lou Y.-Q., 2001, ApJ, 563, L147
%-L150; Magnetohydrodynamic Tidal Waves
%on a Spinning Magnetic Compact Star

\bibitem[\protect\citeauthoryear{Lou \& Cao}{2011}]{b303} Lou Y.-Q.,
Bai X. N., 2011, MNRAS, 415, 925
%--943; ``3D Perturbations in an Isothermal Self-Similar Flow"

\bibitem[\protect\citeauthoryear{Lou \& Cao}{2008}]{b3} Lou Y.-Q.,
Cao Y., 2008, MNRAS, 384, 611
%Self-similar dynamics of a relativistically hot gas

\bibitem[\protect\citeauthoryear{Lou \& Wang}{2006}]{b23} Lou Y.-Q.,
Wang W.-G., 2006, MNRAS, 372, 885

\bibitem[\protect\citeauthoryear{Lou \& Wang}{2007}]{b24} Lou Y.-Q.,
Wang W.-G., 2007, MNRAS, 378, L54

\bibitem[\protect\citeauthoryear{Lou \& Wang}{2011}]{b25} Lou Y.-Q.,
Wang L. L., 2011, MNRAS, in press (2011arXiv1109.2682L)

\bibitem[\protect\citeauthoryear{Murphy, Burrows \& Heger}{2004}]{b11}
Murphy J. W., Burrows A., Heger A., 2004, ApJ, 615, 460

\bibitem[\protect\citeauthoryear{Nordhaus et al.}{2010}]{Nordhaus-PRD}
Nordhaus J., Brandt T. D., Burrows A., Livne E., Ott C. D., 2010a,
Phys. Rev. D, 82, 103016
%Theoretical support for the hydrodynamic mechanism of pulsar kicks
%
%The collapse of a massive star’s core, followed by a neutrino-driven,
%asymmetric supernova explosion, can naturally lead to pulsar recoils
%and neutron star kicks. Here, we present a two-dimensional,
%radiation-hydrodynamic simulation in which core collapse leads
%to significant acceleration of a fully formed, nascent neutron
%star via an induced, neutrino-driven explosion. During the explosion,
%an ?10% anisotropy in the low-mass, high-velocity ejecta leads to
%recoil of the high-mass neutron star. At the end of our simulation,
%the neutron star has achieved a velocity of ?150kms-1 and is
%accelerating at ?350kms-2, but has yet to reach the ballistic regime.
%The recoil is due almost entirely to hydrodynamical processes, with
%anisotropic neutrino emission contributing less than 2% to the
%overall kick magnitude. Since the observed distribution of neutron
%star kick velocities peaks at ?300-400kms-1, recoil due to
%anisotropic core-collapse supernovae provides a natural,
%nonexotic mechanism with which to obtain neutron star kicks.
%

\bibitem[\protect\citeauthoryear{Nordhaus et al.}{2010}]{Nordhaus-APJ} Nordhaus J.,
 Burrows A., Almgren A., Bell J., 2010b, ApJ, 720, 694

\bibitem[\protect\citeauthoryear{Rantsiou et al.}{2011}]{Rantsiou2011} Rantsiou E., Burrows A.,
 Nordhaus J., Almgren A., 2011, ApJ, 732, 57
%The monopole approximation for self-gravity was adopted?!.

\bibitem{bRossby1}Rossby C.-G. 1938, J. Marine Res., 2, 239

\bibitem{bRossby2}Rossby C.-G., et al. 1939, J. Marine Res., 2, 38

\bibitem{SaenShapiro78} Saen R. A., Shapiro S. L.,
1978, ApJ, 221, 286
%--303; Gravitational radiation from
%stellar collapse - Ellipsoidal models

\bibitem[\protect\citeauthoryear{Scheck et al.}{2004}]{Scheck2004}
Scheck L., Plewa T., Janka H. Th., Kifonidis H., Mueller E., 2004,
Phys. Rev. Lett., 92, 011103
%Parts of the work Scheck et al. (2006) already have been
%presented in this Letter (Scheck et al. 2004), but a detailed
%description of both their methods and results are given in
%Scheck et al. (2006).
%

\bibitem[\protect\citeauthoryear{Scheck et al.}{2006}]{Scheck2006} Scheck L.,
Kifonidis H., Janka H. Th., Mueller E., 2006, A\&A, 457, 963
%-986; "Multidimensional supernova simulations with approximative
%neutrino transport I. Neutron star kicks and the anisotropy of
%neutrino-driven explosions in two spatial dimensions"
%
%section 3.1, Our calculations are started at \sim 15-20 ms
%after core core bounce from detailed post-collapse models.
%
%section 4.2, discussion on g-modes of Burrows et al. (2006);
%It is possible, however, that our simulations underestimate
%such effects, which would require the inclusion of the
%whole neutron star without excising the centra core, and the
%ability to follow the excitatio of deep modes due to a
%self-consistent coupling between accretion, core motion, and
%core-mode generation.
%

\bibitem{Dynamo2} Thompson C., Duncan R. C., 1993, APJ, 408, 194
%Neutron Star Dynamos and the Origin of Pulser Magnetism

\bibitem{b46} Unno W., Osaki Y., Ando H., Shibahashi H.,
1979, Nonradial oscillations of stars, University of Tokyo Press,
Tokyo

\bibitem[\protect\citeauthoryear{Van Riper}{1982}]{b03} Van Riper
K. A., 1982, ApJ, 257, 793

\bibitem[\protect\citeauthoryear{Van Riper \& Lattimer}{1981}]{b02}Van
Riper K. A., Lattimer J. M., 1981, ApJ, 249, 270

\bibitem{b40} Wang W.-G., Lou Y.-Q., 2007, Ap\&SS, 311, 363
% Self-similar dynamics of a magnetized polytropic gas

\bibitem{WangLou08} Wang W.-G., Lou Y.-Q., 2008, Ap\&SS, 315, 135
%-156; (arXiv:0804.2889)
%``Dynamic Evolution of a Quasi-Spherical General
%Polytropic Magnetofluid with Self-Gravity"

\bibitem[\protect\citeauthoryear{Wilkinson}{1965}]{b6} Wilkinson J. H.,
1965, The Algebraic Eigenvalue Problem. Clarendon Press, Oxford

\bibitem[\protect\citeauthoryear{Woosley, Langer \& Weaver}{1993}]{b19}
Woosley S. E., Langer N., Weaver T. A., 1993, ApJ, 411, 823

\bibitem[\protect\citeauthoryear{Woosley, Heger \& Weaver}{2002}]{b20}
Woosley S. E., Heger A., Weaver T. A., 2002, Rev. Mod. Phys, 74, 1015

\bibitem[\protect\citeauthoryear{Yahil}{1983}]{b16} Yahil A., 1983,
ApJ, 265, 1047

\bibitem[\protect\citeauthoryear{Yahil&Lattimer}{1982}]{b16b}
Yahil A., Lattimer J. M., 1982, in Supernovae: A Survey of
  Current Research, edited by M. J. Rees and R. S. Stoneham
  (Reidel, Dordrecht), p. 53

\end{thebibliography}
\end{document}